\documentclass[onecolumn,traditabstract,longauth]{aa}

\usepackage{graphicx}
\usepackage{txfonts}
\usepackage{xcolor}

\newcommand{\xmm}{{\em XMM-Newton}}
\newcommand{\Planck}{{\em Planck}}
\graphicspath{{./}{figures/}}
\usepackage[normalem]{ulem}
\begin{document} 

   \title{\mbox{CHEX-MATE: CLUster Multi-Probes in Three Dimensions (CLUMP-3D)}}
   \subtitle{I. Gas Analysis Method using X-ray and Sunyaev--Zel'dovich Effect Data}

   \author{Junhan Kim\inst{1,2}
          \and
          Jack Sayers\inst{1}
          \and
          Mauro Sereno\inst{3,4}
          \and
          Iacopo Bartalucci\inst{5}
          \and
          Loris Chappuis\inst{6}
          \and
          Sabrina De Grandi\inst{7}
          \and
          \mbox{Federico De Luca}\inst{8,9}
          \and
          Marco De Petris\inst{10}
          \and
          Megan E. Donahue\inst{11}
          \and
          Dominique Eckert\inst{6}
          \and
          Stefano Ettori\inst{3,4}
          \and
          \mbox{Massimo Gaspari}\inst{12}
          \and
          Fabio Gastaldello\inst{5}
          \and
          Raphael Gavazzi\inst{13,14}
          \and
          Adriana Gavidia\inst{1}
          \and
          Simona Ghizzardi\inst{5}
          \and
          Asif Iqbal\inst{15}
          \and
          Scott Kay\inst{16}
          \and
          Lorenzo Lovisari\inst{4, 17}
          \and
          Ben J. Maughan\inst{18}
          \and
          Pasquale Mazzotta\inst{8,9}
          \and
          Nobuhiro Okabe\inst{19,20,21}
          \and
          \mbox{Etienne Pointecouteau}\inst{22}
          \and
          Gabriel W. Pratt\inst{15}
          \and
          Mariachiara Rossetti\inst{5}
          \and
          Keiichi Umetsu\inst{23}
          }

   \institute{California Institute of Technology, 
              1200 E. California Blvd., MC 367-17, Pasadena, CA 91125, USA\\
              \email{junhan@kaist.ac.kr}
              \and
              Department of Physics, Korea Advanced Institute of Science and Technology (KAIST), 291 Daehak-ro, Yuseong-gu, Daejeon 34141, Republic of Korea
              \and
              INAF, Osservatorio di Astrofisica e Scienza dello Spazio, via Piero Gobetti 93/3, 40129 Bologna, Italy 
              \and
              INFN, Sezione di Bologna, viale Berti Pichat 6/2, 40127 Bologna, Italy
              \and
              INAF, Istituto di Astrofisica Spaziale e Fisica Cosmica di Milano, via A. Corti 12, 20133, Milano, Italy
              \and
              Department of Astronomy, University of Geneva, Ch. d’Ecogia 16, CH-1290 Versoix, Switzerland
              \and
              INAF - Osservatorio Astronomico di Brera, via E. Bianchi 46, 23807 Merate (LC), Italy
              \and
              Universit\`a degli studi di Roma ‘Tor Vergata’, Via della ricerca scientifica 1, 00133 Roma, Italy
              \and
              INFN, Sezione di Roma ‘Tor Vergata’, Via della Ricerca Scientifica 1, 00133 Roma, Italy
              \and
              Dipartimento di Fisica, Sapienza Universit\`a di Roma, Piazzale Aldo Moro 5, I-00185, Roma, Italy
              \and
              Department of Physics and Astronomy, Michigan State University, 567 Wilson Road, East Lansing, MI 48864, USA
              \and
              Department of Astrophysical Sciences, Princeton University, Princeton, NJ 08544, USA
              \and
              Laboratoire d’Astrophysique de Marseille, Aix-Marseille Univ, CNRS, CNES, Marseille, France
              \and
              Institut d’Astrophysique de Paris, UMR 7095, CNRS \& Sorbonne Universit\'e, 98 bis boulevard Arago, F-75014 Paris, France
              \and
              Universit\'e Paris-Saclay, Universit\'e Paris Cit\'e, CEA, CNRS, AIM, 91191, Gif-sur-Yvette, France 
              \and
              Jodrell Bank Centre for Astrophysics, Department of Physics and Astronomy, The University of Manchester, Manchester M13 9PL, UK
              \and
              Center for Astrophysics $|$ Harvard $\&$ Smithsonian, 60 Garden Street, Cambridge, MA 02138, USA
              \and
              HH Wills Physics Laboratory, University of Bristol, Bristol, BS8 1TL, UK
              \and
              Physics Program, Graduate School of Advanced Science and Engineering, Hiroshima University, Hiroshima 739-8526, Japan
              \and
              Hiroshima Astrophysical Science Center, Hiroshima University, 1-3-1 Kagamiyama, Higashi-Hiroshima, Hiroshima 739-8526, Japan
              \and
              Core Research for Energetic Universe, Hiroshima University, 1-3-1, Kagamiyama, Higashi-Hiroshima, Hiroshima 739-8526, Japan
              \and
              IRAP, Universit\'e de Toulouse, CNRS, CNES, UT3-UPS, (Toulouse), France
              \and
              Academia Sinica Institute of Astronomy and Astrophysics (ASIAA), No. 1, Section 4, Roosevelt Road, Taipei 10617, Taiwan
             }

   \date{}

  \abstract{
   Galaxy clusters are the products of structure formation through myriad physical processes that affect their growth and evolution throughout cosmic history. As a result, the matter distribution within galaxy clusters, or their shape, is influenced by cosmology and astrophysical processes, in particular the accretion of new material due to gravity.
   We introduce an analysis method for investigating the three-dimensional triaxial shapes of galaxy clusters from the Cluster HEritage project with \xmm\ -- Mass Assembly and Thermodynamics at the Endpoint of structure formation (CHEX-MATE).
   In this paper, the first in a CHEX-MATE triaxial analysis series, we focus on utilizing X-ray data from \xmm\ and Sunyaev-Zel'dovich (SZ) effect maps from \Planck\ and the Atacama Cosmology Telescope to obtain a three-dimensional triaxial description of the intracluster medium (ICM) gas. We present the forward modeling formalism of our technique, which projects a triaxial ellipsoidal model for the gas density and pressure, to be compared directly with the observed two-dimensional distributions in X-rays and the SZ effect. A Markov chain Monte Carlo is used to estimate the posterior distributions of the model parameters. Using mock X-ray and SZ observations of a smooth model, we demonstrate that the method can reliably recover the true parameter values.
   In addition, we applied the analysis to reconstruct the gas shape from the observed data of one CHEX-MATE galaxy cluster, PSZ2 G313.33+61.13 (Abell 1689), to illustrate the technique. The inferred parameters are in agreement with previous analyses for the cluster, and our results indicate that the geometrical properties, including the axial ratios of the ICM distribution, are constrained to within a few percent. With a much better precision than previous studies, we thus further establish that Abell 1689 is significantly elongated along the line of sight, resulting in its exceptional gravitational lensing properties.
   }


   \maketitle

%

\section{Introduction} \label{sec:intro}

Galaxy clusters are useful probes of structure formation, astrophysical processes such as shocks and feedback from active galactic nuclei, and cosmology \citep{1985ApJ...292..371D, 2005RvMP...77..207V, 2011ARA&A..49..409A, 2012ARA&A..50..353K,Markevitch2007,McNamara2007}. For instance, they are fundamental to the science goals of numerous ongoing and upcoming large survey projects, including those carried out by \textit{eROSITA} \citep[extended ROentgen Survey with an Imaging Telescope Array;][]{2021A&A...647A...1P}, \textit{Euclid} \citep{2019A&A...627A..23E}, and the \textit{Rubin Observatory} \citep{2019ApJ...873..111I}. In order to maximize the scientific reach of such programs, particularly with regard to cosmological parameter constraints, it is crucial to accurately characterize the ensemble average physical properties of galaxy clusters along with the intrinsic scatter relative to these averages \citep[e.g.,][]{Lima2005, Zhan2018, 2019A&A...627A..23E}. One such example are the scaling relations used to connect global galaxy cluster observables to the underlying halo mass \citep{Rozo2014, Mantz2016, Pratt2019}. While these scaling relations are generally sensitive to a range of astrophysical processes \citep[e.g.,][]{Ansarifard2020}, some observables, including the gravitational weak lensing measurements often used to determine the absolute mass, have deviations from average relations that are dominated by projection effects related to asphericity and orientation \citep{Meneghetti2010, 2011ApJ...740...25B}.

The Cluster HEritage project with \xmm\ -- Mass Assembly and Thermodynamics at the Endpoint of structure formation \citep[CHEX-MATE;][]{2021A&A...650A.104C}\footnote{\url{http://xmm-heritage.oas.inaf.it/}} is an effort to provide a more accurate understanding of the population of galaxy clusters at low-$z$ and with high masses, particularly in the context of cosmology and mass calibration, including the shape of their matter distributions and the effects of the baryonic physics on their global properties. The project is based on a 3~Msec \xmm\ program that observed 118 galaxy clusters, from two equal-sized subsamples selected from the \Planck\ all-sky Sunyaev-Zel'dovich (SZ)\footnote{Throughout this paper, we use the abbreviation `SZ' to represent the thermal Sunyaev-Zel'dovich effect.} effect survey. The CHEX-MATE Tier-1 and Tier-2 samples each include 61 galaxy clusters with four overlapping clusters and represent a volume-limited sample ($0.05 < z \leq 0.2 $) in the local universe and a mass-limited sample ($M_{500} \geq 7.25 \times 10^{14}~\textrm{M}_\odot$)\footnote{The parameter $M_{500}$ denotes the mass enclosed within a radius ($R_{500}$) where the mean overdensity is 500 times the critical density at a specific redshift, and we used the $M_{500}$ and $R_{500}$ values from \cite{2016A&A...594A..27P}.} of the most massive objects in the universe, respectively. The X-ray observing program has recently been completed, and initial results from the analyses of these data along with publicly available SZ data have already been published \citep{2022A&A...665A.117C, 2022arXiv220909601O, 2023AnA...674A.179B}.

We utilized triaxial modeling techniques \citep[e.g.,][]{2013SSRv..177..155L} to investigate the three-dimensional mass distribution within the CHEX-MATE galaxy clusters to infer their intrinsic properties. This approach was motivated by two factors: (1) Three-dimensional triaxial shapes provide a better approximation of galaxy clusters than spherical models, and the parameters obtained from such an analysis, such as mass, have lower levels of bias and intrinsic scatter \citep{2011ApJ...740...25B, 2016MNRAS.463..655K}. (2) A correlation between the triaxial shape of the dark matter (DM) halo and its formation history has been established in simulations \citep[e.g.,][]{2006ApJ...647....8H, 2021MNRAS.500.1029L, 2022A&A...663A..17S}, suggesting that triaxial shape measurements can be a powerful probe of cosmology independent of other techniques currently in use. For instance, some lensing-based shape measurements have found good agreement with $\Lambda$CDM predictions \citep{2010MNRAS.405.2215O, 2018ApJ...860..126C}, while a recent multi-probe triaxial analysis suggests a $\simeq 2\sigma$ discrepancy between the observed and predicted minor to major axial ratio distributions \citep{2018ApJ...860L...4S}.  This discrepancy could indicate that clusters formed more recently than predicted. Alternatively, elevated merger rates \citep{2017MNRAS.466..181D}, a reduced influence of baryons on the DM \citep{2017PASJ...69...14S}, or an enhanced feedback \citep{2004ApJ...611L..73K} could also explain the observed cluster shapes. CHEX-MATE offers a uniform selection of galaxy clusters with consistent measurements of the intracluster medium (ICM) density and temperature. This clean, well-characterized selection with a large sample size ($\sim\!80$ clusters excluding major mergers; \citealt{2022A&A...665A.117C}) will enable a robust cosmological measurement of the triaxial shape distribution.

For our analysis, we adopted the CLUster Multi-Probes in Three Dimensions \citep[CLUMP-3D;][]{2017MNRAS.467.3801S, 2018ApJ...860L...4S, 2018ApJ...860..126C, 2021MNRAS.505.4338S} project and implemented significant updates to the modeling package. CLUMP-3D incorporates multiwavelength data from X-ray observations (surface brightness and spectroscopic temperature), millimeter-wave observations (SZ surface brightness), and optical observations (gravitational lensing), which are the projected observables. Then, it assumes triaxial distributions of the ICM gas and matter density profiles. Taking advantage of the different dependences of the X-ray and SZ signals on the gas density and temperature, these signals are used to probe the line-of-sight extent of the ICM, and gravitational lensing data are used to probe the projected matter distribution. In particular, the X-ray emission observed from the ICM is proportional to the line-of-sight integral of the squared electron density ($n_e$) multiplied by the X-ray cooling function, $\Lambda$, represented as $S_X \propto \int n^2_e \Lambda dl$. Meanwhile, the detected SZ signal is proportional to the line-of-sight integral of the product of electron density and temperature ($T_e$), denoted as $B_\textrm{SZ} \propto \int n_e T_e dl$. Given that the ICM temperature ($T_X$) can be spectroscopically measured using X-ray observations, the line-of-sight elongation ($\triangle l$) can subsequently be determined through the combination of these three measurements as $\triangle l \sim (B^2_\textrm{SZ} \Lambda) / (S_X T^2_X)$. Assuming a co-alignment of the triaxial axes of the ICM and DM distributions, while still allowing for different axial ratios for the two quantities, our multi-probe analysis can thus constrain the three-dimensional shapes of galaxy clusters. CLUMP-3D was introduced in \cite{2017MNRAS.467.3801S}, where the authors inferred the triaxial matter and gas distribution of the galaxy cluster MACS J1206.2$-$0847. The technique built upon similar methods developed to constrain cluster morphology \citep[e.g.,][]{2011MNRAS.416.3187S, 2012MNRAS.419.2646S}. It was then applied to measure the shapes of the Cluster Lensing And Supernova survey with Hubble \citep[CLASH\footnote{\url{https://www.stsci.edu/~postman/CLASH/}};][]{2012ApJS..199...25P} clusters, to probe the ensemble-averaged three-dimensional geometry \citep{2018ApJ...860L...4S, 2018ApJ...860..126C} as well as the radial profile of the nonthermal pressure fraction \citep{2021MNRAS.505.4338S}. These results demonstrate the potential of the three-dimensional triaxial shape measurement technique, but they were relatively imprecise due to the sample size, data quality, and systematics related to cluster selection. Thus, the much larger CHEX-MATE galaxy cluster sample, with a well-understood selection function and more uniform and higher-quality X-ray data, will provide improved statistics and more robust constraints on the shape measurements.

In this paper we demonstrate several improvements to the original CLUMP-3D formalism while modeling the ICM distributions observed by \xmm and \Planck and ground-based SZ data from the Atacama Cosmology Telescope (ACT). As detailed below, we performed a fully two-dimensional analysis of the X-ray temperature \citep{2023arXiv231102176L} and SZ data, whereas the original CLUMP-3D only treated the X-ray surface brightness (SB) in two dimensions while using one-dimensional azimuthally averaged profiles of both the X-ray spectroscopic temperatures and the SZ effect data. In addition, we now model the ICM gas density and pressure instead of its density and temperature. This allows us to fit the data with fewer parameters, thus accelerating the model fitting process. Additionally, we fully rewrote the code in Python to facilitate the future public release of the package. In Sect.~\ref{sec:formalism} we summarize the triaxial analysis formalism and describe the model fitting method. In Sect.~\ref{sec:application_chexmate} we introduce the X-ray and SZ data from our program and apply the technique to a CHEX-MATE galaxy cluster. In a subsequent paper, we will include gravitational lensing constraints in a manner that also builds upon, and improves, the existing CLUMP-3D technique. With these X-ray, SZ effect, and gravitational lensing data, we will be able to model the triaxial distributions of both the ICM and the DM. Throughout this study, we adopt a $\Lambda$CDM cosmology characterized by $H_0 = 70~\textrm{km}~\textrm{s}^{-1}~\textrm{Mpc}^{-1}$, $\Omega_\textrm{m} = 0.3$, and $\Omega_\Lambda = 0.7$. $E(z)$ represents the ratio of the Hubble constant at redshift, $z$, to its present value, $H_0$, and $h_{70} = H_0 / (100~\textrm{s}^{-1}~\textrm{Mpc}^{-1})/0.7$.

\section{Triaxial analysis: Formalism and the model fit}\label{sec:formalism}

While the mathematical description of a triaxial geometry for astronomical objects and their physical profiles has been introduced in previous studies \citep[e.g.,][]{1977ApJ...213..368S, 1980A&A....82..289B, 1985MNRAS.212..767B, 2003ApJ...599....7O, 2005ApJ...625..108D, 2007MNRAS.380..149C, 2010MNRAS.403.2077S, 2012MNRAS.419.2646S, 2018ApJ...860L...4S}, these works lack consistency in their notation. To prevent confusion, we present our mathematical formalism for triaxial modeling in this section. We then describe our model fitting procedure and the implementation of the fitting algorithm in our software package.

As this paper focuses on the analysis method of studying ICM distributions, we do not include the gravitational lensing data in our fits. Future works in this series will expand our formalism to include total mass density profiles constrained by gravitational lensing measurements. For instance, in the case of a Navarro–Frenk–White \citep[NFW;][]{1997ApJ...490..493N} profile the gravitational lensing analysis requires two additional parameters -- total mass ($M_{200}$) and concentration ($c_{200}$) -- assuming that the gas and matter distributions are co-aligned along the ellipsoidal axes. This assumption is well supported by a two-dimensional weak-lensing and X-ray analysis of 20 high-mass galaxy clusters \citep{2018ApJ...860..104U}, as well as by cosmological hydrodynamical simulations \citep{2018MNRAS.478.1141O}.

\subsection{Geometry and projection}\label{subsec:geom}

To connect the intrinsic cluster geometry to the projected properties observed in the plane of the sky, we assume a triaxial ellipsoidal model for the gas distribution, where the thermodynamic profiles of the ICM are represented as a function of $\zeta$, the ellipsoidal radius. In the intrinsic coordinate system of the ellipsoid $(x_1, x_2, x_3)$, it is defined as:
\begin{equation}
    \zeta^2 = \frac{x^2_1}{q^2_1} + \frac{x^2_2}{q^2_2} + x^2_3,
    \label{eq:ellipsoid3d}
\end{equation}
where $q_1$ and $q_2$ are minor-to-major and intermediate-to-major axial ratios, respectively ($0 < q_1 \leq q_2 \leq 1$). Given a semimajor axis of the ellipsoid $l_s$, the volume of the ellipsoid is $(4 \pi / 3) {l^3_s} q_1 q_2$. The ellipsoid becomes a prolate shape if $q_1 = q_2 \leq 1$ and an oblate shape if $q_1 \leq q_2 = 1$.

\begin{figure}[t!]
    \centering
    \includegraphics[width=0.5\textwidth]{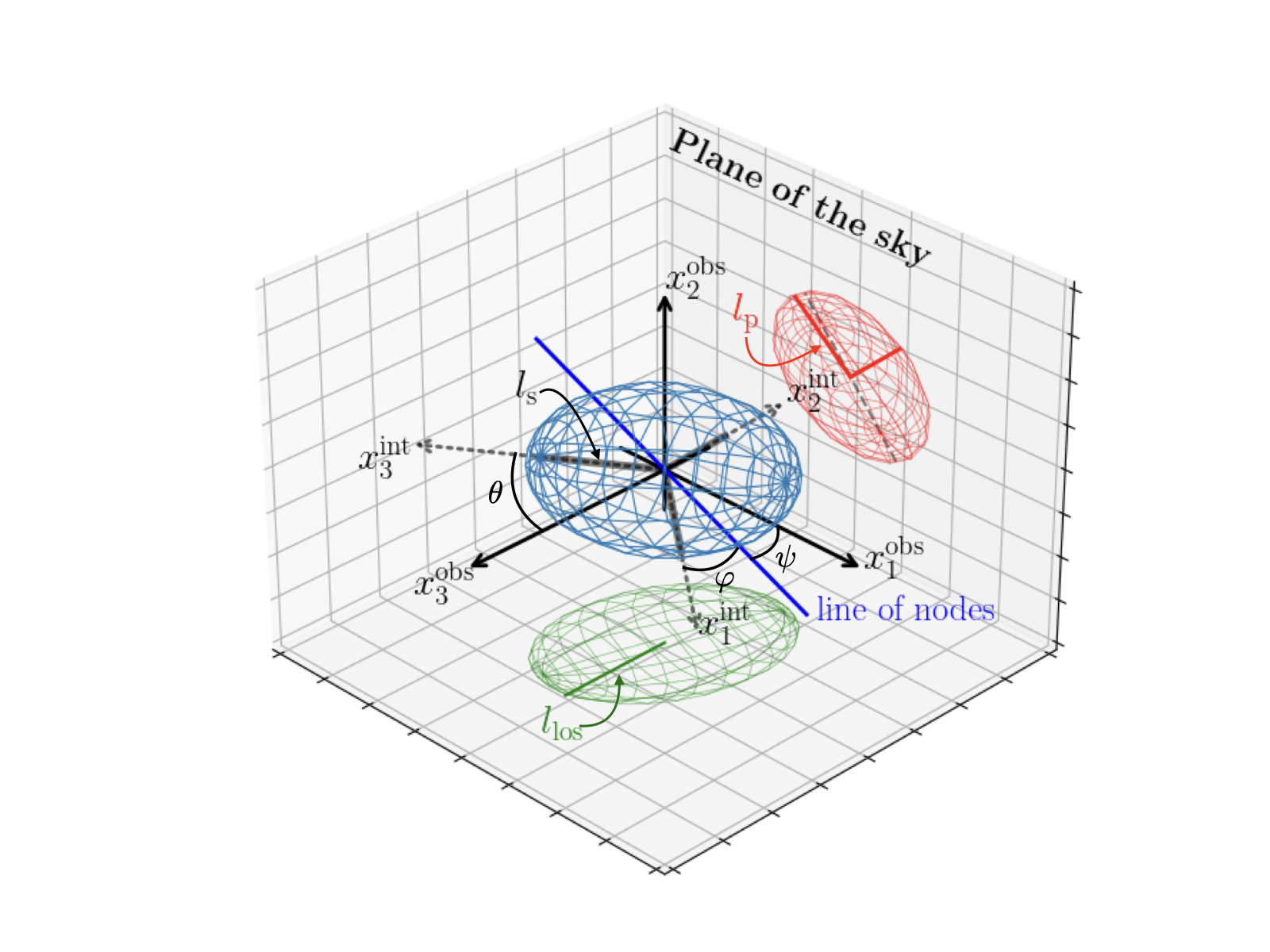}
    \caption{Triaxial ellipsoid model and coordinate systems used in the triaxial analysis. The intrinsic coordinate system of the ellipsoid is denoted by dotted gray arrows ($x^{\text{\tiny int}}_1$, $x^{\text{\tiny int}}_2$, and $x^{\text{\tiny int}}_3$), where $x^{\text{\tiny int}}_3$ represents the major axis. The black arrows ($x^{\text{\tiny obs}}_1$, $x^{\text{\tiny obs}}_2$, and $x^{\text{\tiny obs}}_3$) correspond to the observer's coordinate system, where $x^{\text{\tiny obs}}_3$ is aligned with the observer's line of sight. In other words, an observer views the ellipsoid in the $-x^{\text{\tiny obs}}_3$ direction. The three Euler angles ($\theta$, $\varphi$, $\psi$) characterize the intrinsic coordinate system of the ellipsoid in relation to the observer's coordinate system. The blue line represents the line of nodes, which is the intersection of the $x^{\text{\tiny int}}_1$-$x^{\text{\tiny int}}_2$ plane and the $x^{\text{\tiny obs}}_1$-$x^{\text{\tiny obs}}_2$ plane, and it is aligned with the vector $x^{\text{\tiny int}}_3 \times x^{\text{\tiny obs}}_3$. The red ellipse denotes the projection of the ellipsoid on the sky plane, with $l_\textrm{p}$ representing its semimajor axis. The dashed black line on the ellipse shows the projected major axis of the ellipsoid on the sky plane. The green ellipse is the projection of the ellipsoid onto the plane that is perpendicular to the sky plane, and $l_{\text{\tiny los}}$ is the half size of the ellipse along the observer line of sight. See also Figs.~2 and 3 in \cite{2012MNRAS.419.2646S}.
              }
    \label{fig:ellipsoid}
\end{figure}

Figure~\ref{fig:ellipsoid} illustrates the geometry of the ellipsoid and the involved coordinate systems. It is essential to note that the axes defining the ICM model may not align with the observer's frame. To relate the ellipsoid's intrinsic coordinate system ($x^{\text{\tiny int}}_1$, $x^{\text{\tiny int}}_2$, $x^{\text{\tiny int}}_3$) to the observer's coordinate frame ($x^{\text{\tiny obs}}_1$, $x^{\text{\tiny obs}}_2$, $x^{\text{\tiny obs}}_3$), we employ three Euler angles. These angles describe the relationship between the two coordinate systems: (1) the angle between $x^{\text{\tiny int}}_3$, aligned with the major axis of the ellipsoid, and $x^{\text{\tiny obs}}_3$, which lies along the observer's line of sight ($\theta$), (2) the angle between $x^{\text{\tiny int}}_1$ and the line of nodes ($\varphi$), and (3) the angle between $x^{\text{\tiny obs}}_1$ and the line of nodes ($\psi$). The line of nodes is the intersection of the $x^{\text{\tiny int}}_1$-$x^{\text{\tiny int}}_2$ plane and the $x^{\text{\tiny obs}}_1$-$x^{\text{\tiny obs}}_2$ plane, and it is aligned with the vector $x^{\text{\tiny int}}_3 \times x^{\text{\tiny obs}}_3$.

We can derive the geometric properties of the projected ellipse from the intrinsic parameters of the ellipsoid when it is projected onto the plane from any direction. These properties encompass the semimajor axis of the projected ellipse  $l_\textrm{p}$, its ellipticity $\epsilon$, the orientation of the ellipse in the plane of the sky $\theta_\epsilon$, and the elongation parameter $e_\parallel$. The projected profiles are expressed as a function of $\xi$, the elliptical radius of the ellipse in the plane of the sky.

The ellipticity of the projected ellipse ($\epsilon$) is
\begin{equation}
    \epsilon = 1-q_\textrm{p},
\end{equation}
where $q_\textrm{p}$ is the minor-to-major axial ratio of the observed projected isophote ($q_\textrm{p} \leq 1$), which is the inverse of $e_\textrm{p}$ used in \cite{2012MNRAS.419.2646S}, and
\begin{equation}
    q_\textrm{p} = \sqrt{\frac{j + l - \sqrt{(j - l)^2 + 4k^2}}{j + l + \sqrt{(j - l)^2 + 4k^2}}},
\end{equation}
where
\begin{equation}
\begin{split}
    j &= \frac{1}{2}\Biggl[\left(\frac{1}{q^2_1} + \frac{1}{q^2_2} \right) - \frac{\sin^2\theta \cos^2 \psi \left(q^2_1 + q^2_2 - 2 \right)}{q^2_1 q^2_2} + \\
    & \left(\frac{1}{q^2_1} - \frac{1}{q^2_2} \right) \left\{ \cos 2\varphi \left( \cos^2\theta \cos^2\psi - \sin^2\psi \right) - \cos\theta \sin2\varphi \sin2\psi \right\} \Biggr],\nonumber
\end{split}
\end{equation}

\begin{equation}
\begin{split}
    k &= \frac{1}{4 q^2_1 q^2_2}\biggl[2 \cos\theta \left(q^2_1 - q^2_2 \right) \cos 2\psi \sin 2\varphi + \\ 
    & \left\{\sin^2\theta \left(q^2_1 + q^2_2 - 2 \right) + \left(1+\cos^2\theta \right) \left(q^2_1 - q^2_2 \right) \cos 2\varphi \right\} \sin2\psi \biggr],\nonumber
\end{split}
\end{equation}

\begin{equation}
\label{eq:fproj}
\begin{split}
    l &= \frac{1}{2}\Biggl[\left(\frac{1}{q^2_1} + \frac{1}{q^2_2} \right) - \frac{\sin^2\theta \sin^2 \psi \left(q^2_1 + q^2_2 - 2 \right)}{q^2_1 q^2_2} + \\
    & \left(\frac{1}{q^2_1} - \frac{1}{q^2_2} \right) \left\{ \cos 2\varphi \left( \cos^2\theta \sin^2\psi - \cos^2\psi \right) + \cos\theta \sin2\varphi \sin2\psi \right\} \Biggr].
\end{split}
\end{equation}
    
It is worth noting that the expressions of $j$, $k$, and $l$ in \cite{1977ApJ...213..368S} and \cite{1980A&A....82..289B} differ from those presented above, as they assumed $\psi = 0$, using only two angles to align the major ellipsoidal axis with the observer's line of sight. However, a coordinate transformation requiring $\psi$ is necessary to align the remaining axes.

The orientation angle in the plane of the sky of the projected ellipse is
\begin{equation}
    \theta_\epsilon = \tan^{-1} \left(\frac{l-j + \sqrt{(j - l)^2 + 4k^2}}{2k}\right),
\end{equation}
and the elongation parameter of the ellipsoid is 
\begin{equation}\label{eq:elongation}
    e_\parallel \equiv \frac{l_\textrm{los}}{l_\textrm{p}} =  \sqrt{\frac{q_\textrm{p}}{q_1 q_2}} f^{-3/4},
\end{equation}
where
\begin{equation}
    f = \sin^2 \theta \left[ \left( \frac{\sin \varphi}{q_1} \right)^2 + \left( \frac{\cos \varphi}{q_2} \right)^2 \right] + \cos^2 \theta.
\end{equation}
The elongation parameter, $e_\parallel$, represents the ratio of the size of the ellipsoid along the observer's line of sight to the major axis of the projected ellipse in the sky plane, providing a measure of the three-dimensional geometry of the triaxial ellipsoid model of the ICM. In the gas analysis presented in \cite{2012MNRAS.419.2646S}, the orientation angle ($\theta_\epsilon$; represented as $\epsilon$ by \citealt{2012MNRAS.419.2646S}) was determined from the X-ray map, while the elongation parameter (represented as $e_\triangle$, which is equivalent to $1 / e_\parallel$) was estimated from the combined X-ray and SZ analysis. Later, \cite{2017MNRAS.467.3801S} simultaneously constrained the individual Euler angles by treating the axial ratios and three angles as free parameters.



Then, the semimajor axis of the projected ellipse becomes
\begin{eqnarray}
    l_\textrm{p} &=& \frac{l_s}{e_\parallel \sqrt{f}}\\
    &=& l_s \sqrt{\frac{q_1 q_2}{q_\textrm{p}}} f^{1/4},
\end{eqnarray}
and the projected length scales $l_s$ and $l_\textrm{los}$ are related by the elongation parameter, that is,
\begin{equation}
    l_\textrm{los} = l_s / \sqrt{f}.
\end{equation}
In the plane of the sky, an elliptical radius $\xi$ becomes
\begin{equation}
    \xi^2 = \left(x^2_1 + \frac{x^2_2}{q^2_\textrm{p}} \right) \left(\frac{l_s}{l_\textrm{p}} \right)^2
\end{equation}
\citep{2010MNRAS.403.2077S}.
\footnote{Assuming that the ellipse is expressed as $\frac{x^2_1}{a^2}+ \frac{x^2_2}{b^2} = 1$, $q_\textrm{p}$ is the minor-to-major axial ratio ($b/a$), and the elliptical radius, which is the corresponding major axis length, becomes $\sqrt{x^2_1 + \frac{x^2_2}{q^2_\textrm{p}}}$ because $x^2_1 + \frac{a^2}{b^2} x^2_2 = a^2$. }

Finally, three-dimensional volume density can be projected onto the sky plane by utilizing the geometric parameters
\begin{eqnarray}
    \label{eq:proj}
    F_{\text{\tiny 2D}} (\xi; l_\textrm{p}, p_i) &=& \frac{2}{\sqrt{f}} \int_{\xi}^{\infty} F_{\text{\tiny 3D}} (\zeta; l_s, p_i) \frac{\zeta}{\sqrt{\zeta^2 - \xi^2}} d\zeta,\\
    F_{\text{\tiny 2D}} (x_\xi; l_\textrm{p}, p_i) &=& 2 l_\textrm{p} e_\parallel \int_{x_\xi}^{\infty} F_{\text{\tiny 3D}} (x_\zeta; l_s, p_i) \frac{x_\zeta}{\sqrt{x^2_\zeta - x^2_\xi}} dx_\zeta,
\end{eqnarray}
where $x_\zeta = \zeta / l_s$, $x_\xi = \xi/l_\textrm{p}$, and $p_i$ are the parameters describing the intrinsic density profile \citep{1977ApJ...213..368S, 2007MNRAS.380.1207S, 2010MNRAS.403.2077S}. Using this projection, we calculated the SZ and X-ray maps on the sky plane from the three-dimensional ellipsoidal distribution of the ICM profiles and fit the model to the data. We describe the analytic profiles ($F_{\text{\tiny 3D}}$) for the physical quantities related to the direct observables ($F_{\text{\tiny 2D}}$) in the next section.

\subsection{Electron density and pressure profiles}

We used smooth analytic functions of the electron density and pressure profiles to describe the thermodynamics and spatial distribution of the ICM, and then used these functions to compute observable quantities, such as the SZ effect map, the X-ray SB map, and the X-ray temperature map. The model lacks the ability to effectively constrain small-scale structures that deviate from its assumptions. However, the three-dimensional description of the profiles provides a better approximation compared to spherical models. After accounting for instrumental effects, such as the point spread function (PSF), these model maps are then compared to the observed data. The original CLUMP-3D package, as detailed in \cite{2017MNRAS.467.3801S}, instead assumed smooth analytic functions for the gas density and temperature \citep{2006ApJ...640..691V, 2012A&A...545A..41B}. However, because the presence (or not) of a cool core alters the overall shape of the temperature profile \citep[e.g.,][]{2007A&A...461...71P}, the analytic function needs to be sufficiently flexible to allow for either a decrease or increase in temperature at small radii. Pressure profiles are more regular in their global shape \citep[e.g.,][]{2010A&A...517A..92A}, and therefore a simpler function with fewer free parameters can be used to describe the ICM. Thus, our overall model can be more easily constrained than the one used by \cite{2017MNRAS.467.3801S}. Table~\ref{tab:params} lists the model parameters used in our gas analysis, including the geometric parameters described in the previous section.

The electron density profile is described as
\begin{equation}
    \label{eq:ne}
    n_e (\zeta) = n_0 \left(\frac{\zeta}{\zeta_c}\right)^{-\eta_e} \left[1 + \left(\frac{\zeta}{\zeta_c}\right)^2 \right]^{-3\beta_e/2+\eta_e/2}\left[1 + \left(\frac{\zeta}{\zeta_t}\right)^3 \right]^{-\gamma_e/3},
\end{equation}
where $n_0$ is the central electron density, $\zeta_c$ is the core radius, and $\zeta_t$ is the tidal radius ($\zeta_t$ $>$ $\zeta_c$). ($\beta_e$, $\eta_e$, $\gamma_e$) represent the power law exponent of the electron density distribution for the intermediate, inner, and external slope of the profile, respectively \citep{2006ApJ...640..691V, 2009A&A...501...61E}. The electron pressure profile is modeled using a generalized NFW (gNFW) profile \citep{1996ApJ...462..563N, 2007ApJ...668....1N, 2010A&A...517A..92A}. It is described as
\begin{equation}
    \label{eq:pressure}
    \frac{P_e(x)}{P_{500}} = \frac{P_0}{(c_{500}x)^{\gamma_p} [1 + (c_{500}x)^{\alpha_p}]^{(\beta_p - \gamma_p)/\alpha_p}},
\end{equation}
where $x = \zeta/R_{500}$, ($\gamma_p$, $\alpha_p$, $\beta_p$) describes the power law exponent for the central ($r \ll r_{\text{\tiny s}}$), intermediate ($r \sim r_{\text{\tiny s}}\!=\!R_{500} / c_{500}$), and outer ($r \gg r_{\text{\tiny s}}$) regions, and the characteristic pressure is
\begin{equation}
    P_{500} = 1.65 \times 10^{-3} E(z)^{8/3} \times \left[ \frac{M_{500}}{3 \times 10^{14} h^{-1}_{70} M_\odot} \right]^{2/3} h^2_{70}~\textrm{keV}~\textrm{cm}^{-3}.
\end{equation}
The expressions for $P_{500}$ provided in \cite{2007ApJ...668....1N} and \cite{2010A&A...517A..92A} represent the gas pressure and the electron pressure, respectively. We opted to use the electron pressure formulation. In order to convert the electron pressure, $P_e$, into gas pressure, it is necessary to incorporate both the mean molecular weight and the mean molecular weight per free electron into the calculations. As noted by \cite{2007ApJ...668....1N}, strong degeneracies between the pressure profile parameters generally prevent meaningful constraints when all are varied \citep[see also][]{2012ApJ...758...75B}. For our baseline fits, we thus fixed the values of $c_{500}$ and $\gamma_p$ to 1.4 and 0.3 as in \citet{2023ApJ...944..221S}. In addition, because $\beta_p$  characterizes the pressure profile in the outer regions, it may not be well constrained depending on the map size chosen for the actual fit. For the demonstration of our approach using actual CHEX-MATE data in Sect.~\ref{sec:application_chexmate}, we restrict the map size of the X-ray and SZ observational data to within $R_{500}$ to mask out potential spurious signal at large radii that do not originate from a target cluster, and therefore an external constraint on the value of $\beta_p$ is required. In such cases, we used a value that depends on the mass and redshift, given by
\begin{equation}
    \label{eq:sayers22}
    \beta_p = 5.495 \left( \frac{M_{500}}{10^{15}~\textrm{M}_\odot} \right)^{0.15} \left(1 + z \right)^{0.02}.
\end{equation}
This relation is derived from a combined X-ray and SZ analysis of galaxy clusters with a redshift range of $0.05 \leq z \leq 0.60$ and mass range of $4 \times 10^{14} \leq M_{500} \leq 30 \times 10^{14} \textrm{M}_\odot$ \citep{2023ApJ...944..221S}. This fit is thus valid for the mass and redshift ranges of the CHEX-MATE clusters, with Tier-1 covering $0.05<z<0.2$ and $2\times10^{14} < M_{500} < 9\times10^{14} \textrm{M}_\odot$, and Tier-2 encompassing $z<0.6$ and $M_{500} > 7.25\times10^{14} \textrm{M}_\odot$.


\begin{table*}[!t]
\caption{\label{tab:params} Gas model parameters}
    \centering
    \begin{tabular}{ccp{0.5\textwidth}l}
    \hline\hline
    Parameter & Units & Description & Default Prior \\
    \hline
    \multicolumn{4}{c}{Geometrical Parameters of a Triaxial Ellipsoid (Eqs.~\ref{eq:ellipsoid3d} and \ref{eq:fproj})}\\
    \hline
    $q_{\text{\tiny ICM,1}}$ &  & Minor-to-major axial ratio of the ICM distribution & $\mathcal{U} (0, 1)$ \\
    $q_{\text{\tiny ICM,2}}$ &  & Intermediate-to-major axial ratio of the ICM distribution & $\mathcal{U} (q_{\text{\tiny ICM,1}}, 1)$ \\
    $\cos \theta$ &  & Cosine of the inclination angle of the ellipsoid major axis & $\mathcal{U} (0, 1)$ \\
    $\varphi$ & deg & Second Euler angle & $\mathcal{U}$ (-$\pi/2$, $\pi/2$) \\
    $\psi$ & deg & Third Euler angle & $\mathcal{U}$ (-$\pi/2$, $\pi/2$) \\ \hline
    \multicolumn{4}{c}{Electron Density Profile (Eq.~\ref{eq:ne})
    } \\ \hline
    $n_0$ & cm$^{-3}$ & Central scale density of the distribution of electrons & $\mathcal{U} (10^{-6}, 10)$ \\
    $\zeta_c$ & kpc & Ellipsoidal core radius of the gas distribution & $\mathcal{U} (0, 10^{3})$ \\
    $\zeta_t$ & Mpc & Ellipsoidal truncation radius of the gas distribution ($\zeta_t > \zeta_c$) & $\mathcal{U}$ ($\zeta_c/10^3$, 3) \\
    $\beta_e$ &  & Slope of the gas distribution (in the intermediate region) & $\mathcal{U} (0, 3)$ \\
    $\eta_e$ &  & Slope of the gas distribution (inner) & $\mathcal{U} (0, 1)$ \\
    $\gamma_e$ &  & Slope of the gas distribution (outer) & $\mathcal{U} (0, 5)$ \\ \hline
    \multicolumn{4}{c}{Gas Pressure Profile (Eq.~\ref{eq:pressure})
    }\\ \hline
    $P_0$ &  & Normalization for the gNFW pressure profile & $\mathcal{U} (0, 10^2)$ \\
    $c_{500}$ &  & Pressure profile concentration ($r \sim r_{\text{\tiny s}} = R_{500} / c_{500}$) & $\delta(1.4)$ \\
    $\gamma_p$ &  & Slope parameter for central region ($r \ll r_{\text{\tiny s}}$)  & $\delta(0.3)$ \\
    $\alpha_p$ &  & Slope parameter for intermediate region ($r \sim r_{\text{\tiny s}}$) & $\mathcal{U} (0, 5)$ \\
    $\beta_p$ &  & Slope parameter for outer region ($r \gg r_{\text{\tiny s}}$) & $\mathcal{U} (0, 15)\tablefootmark{a}$ \\ \hline
    \hline
    \end{tabular}
    \tablefoot{
    We consider five geometric parameters ($q_{\text{\tiny ICM,1}}$, $q_{\text{\tiny ICM,2}}$, $\theta$, $\varphi$, $\psi$), six electron density parameters ($n_0$, $\zeta_\text{\tiny c}$, $\zeta_t$, $\beta_e$, $\eta_e$, $\gamma_e$), and five gas pressure parameters ($P_0$, $c_{500}$, $\gamma_p$, $\alpha_p$, $\beta_p$). For the geometric and electron pressure profile parameters, we primarily adopt the priors in \cite{2018ApJ...860L...4S}. We also assign delta priors to $c_{500}$ (=1.4), and $\gamma_p$ (=0.3) as a default, resulting in 14 free parameters. In the default prior column, $\mathcal{U}$ refers to a uniform prior and $\delta$ refers to a delta function that fixes the parameter for the model fit.\\
    \tablefoottext{a}{For the cluster PSZ2 G313+61.13, to which we applied the model fit in this paper (Sec.~\ref{sec:application_chexmate}), we employed a delta prior (Eq.~\ref{eq:sayers22}) because we limited the map size to be within $R_{500}$, which results} in very little sensitivity to $\beta_p$ \citep{2023ApJ...944..221S}.}
    \vspace{2pt}
\end{table*}

\subsection{Sunyaev-Zel'dovich effect and X-ray observables}\label{subsec:observables}

In this section, we summarize the observables associated with the SZ effect and the X-ray emissivity, and explain their relationship to the electron density and pressure profiles introduced earlier. The SZ effect is characterized by the Compton-$y$ parameter, which is proportional to the integrated electron pressure along the line of sight.
\begin{equation}
    \label{eq:sz}
    y \equiv \frac{\sigma_{\text{\tiny T}}}{m_e c^2} \int_\parallel P_e dl = \frac{\sigma_{\text{\tiny T}} k_{\text{\tiny B}}}{m_e c^2} \int_\parallel n_e T_e dl,   
\end{equation}
where $\sigma_\textrm{T}$ is the Thomson cross-section, $k_B$ is the Boltzmann constant, $n_e$ is the electron number density, and $T_e$ is the electron temperature. The X-ray observations are primarily sensitive to the SB of the ICM,
\begin{equation}
    \label{eq:xray}
    \textrm{SB} = \frac{1}{4 \pi (1+z)^3} \int_\parallel n^2_e \Lambda_{\text{\tiny eff}} (T_e, Z) dl
\end{equation}
\citep{2010ApJ...721..653R}, due to thermal Bremsstrahlung, where the cooling function $\Lambda_{\text{\tiny eff}} (T_e, Z)$ quantifies the thermal radiation emitted from a fully ionized plasma due to collisions, taking into account the relative abundance of each chemical element. It can be calculated using software such as \texttt{XSPEC} \citep{1996ASPC..101...17A}. We used a precalculated table and interpolated the value in the temperature ($T_e$)--metallicity ($Z$) space during the model computation. To calculate the emissivity, the instrument response within the chosen energy band [0.7--1.2] keV and the Galactic hydrogen column density must be taken into account, as explained in \cite{2023AnA...674A.179B}, which describes the details of the data analysis used to produce the SB maps. With our software, we performed the calculation using the Python package \texttt{pyproffit}\footnote{\url{https://pyproffit.readthedocs.io/en/latest/intro.html}} \citep{2020OJAp....3E..12E}.

The \xmm\ data can also be used to derive projected temperature maps of ICM via spectroscopic fits \citep{2023arXiv231102176L}. Within our model, we approximated this spectroscopic temperature based on the formalism of \citet{2004MNRAS.354...10M} as follows:
\begin{equation}
    T_{\text{\tiny sp}} = \frac{\int W T_e dV}{ \int W dV}~\textrm{keV}; ~W = \frac{n^2_e}{T^{3/4}_e},
\end{equation}
which is valid for Bremsstrahlung ($T_e \geq$~3~keV).

The SZ and X-ray observables (Eqs.~\ref{eq:sz} and \ref{eq:xray}) are modeled as projections of the three-dimensional profiles parameterized by the ellipsoidal radius $\zeta$ (or $x_\zeta$). The three-dimensional volume density of the models, $F_{\text{\tiny 3D}} (x_\zeta; l_s, p_i)$, can be written analytically, and the two-dimensional maps are calculated following Eq.~\ref{eq:proj}. The model Compton-$y$ parameter is
\begin{equation}
    y_{\text{\tiny model}} (x_\xi; l_{\text{\tiny p}}, p_i) = \left(2 l_{\text{\tiny p}} e_\parallel \right) \left(\frac{\sigma_{\text{\tiny T}}}{m_e c^2}\right) \int_{x_\xi}^{\infty} P_e (x_\zeta) \frac{x_\zeta}{\sqrt{x^2_\zeta - x^2_\xi}} dx_\zeta,
\end{equation}
where
\begin{equation}
    P_e(x_\zeta) = \frac{P_0 P_{500}}{\left(c_{500}x_\zeta \frac{l_s}{R_{500}} \right)^{\gamma_p} \left[1 + \left(c_{500} x_\zeta \frac{l_s}{R_{500}} \right)^{\alpha_p} \right]^{(\beta_p - \gamma_p)/\alpha_p}}~\textrm{keV cm}^{-3}.
\end{equation}
This integration can be computationally expensive, depending on the size of the map. To expedite the calculation, we created a linearly spaced sample of the (normalized) elliptical radius $x_\xi$ and interpolated the integration results while generating a model map. We applied the same technique in the X-ray observable calculation. Lastly, we convolved the model map with the appropriate PSF shape (e.g., a 7\arcmin\ full width at half maximum Gaussian map in the case of \Planck\ and a 1.6\arcmin\ full width at half maximum in the case of ACT; see Fig.~\ref{fig:paramtest_model}).

Similarly, the X-ray SB (Eq.~\ref{eq:xray}) model becomes
\begin{equation}
\begin{split}
    \textrm{SB}_{\text{\tiny model}} (x_\xi; l_{\text{\tiny p}}, p_i) & = \left( 2 l_{\text{\tiny p}} e_\parallel \right) \frac{1}{4 \pi (1+z)^3} \\
    & \int_{x_\xi}^{\infty} n^2_e (x_\zeta) \Lambda_{\text{\tiny eff}} \left(T_e(x_\zeta), Z (x_\zeta) \right) \frac{x_\zeta}{\sqrt{x^2_\zeta - x^2_\xi}} dx_\zeta,
\end{split}
\end{equation}
where
\begin{equation}
    \label{eq:ne}
    n_e (x_\zeta) = n_0 \left(x_\zeta \frac{l_s}{\zeta_c}\right)^{-\eta_e} \left[1 + \left(x_\zeta \frac{l_s}{\zeta_c}\right)^2 \right]^{-3\beta_e/2+\eta_e/2}\left[1 + \left(x_\zeta \frac{l_s}{\zeta_t}\right)^3 \right]^{-\gamma_e/3},
\end{equation}
and the electron temperature is
\begin{equation}
    T_e(x_\zeta) = \frac{P_e (x_\zeta)}{n_e(x_\zeta) k_B}.
\end{equation}
We used a radius-dependent metallicity profile $Z (x_\zeta)$ obtained from the X-COP galaxy cluster samples \citep{2021A&A...646A..92G} for calculating the cooling function.

Upon generating the model, instrumental responses are incorporated to facilitate a direct comparison between the model and the data. For the \xmm\ X-ray maps, the sky background in the [0.7–1.2] keV band ($2.49 \times 10^{-4}$ cts/s/arcmin$^2$; \citealt{2023AnA...674A.179B}) is considered. Specifically, we adopted the sky and particle background measured by the European Photon Imaging Camera \citep[EPIC;][]{2001A&A...365L..18S, 2001A&A...365L..27T} M2 CCD in the [0.5-2]~keV band and converted it for the [0.7–1.2] keV band. After adding the sky background, the vignetting is applied. Subsequently, the resulting map is convolved with a Gaussian profile to account for the PSF. The nominal PSF of \xmm\ can be can be closely represented using a Gaussian function with a 6\arcsec\ full width at half maximum (FWHM)\footnote{\url{https://xmm-tools.cosmos.esa.int/external/xmm_user_support/documentation/uhb/onaxisxraypsf.html}}. However, the actual FWHM of the PSF is dependent on the angle relative to the optical axis, and combining images from different cameras could potentially deteriorate the final PSF. Therefore, we followed the convention of \cite{2023AnA...674A.179B} and assumed the Gaussian has a FWHM of 10\arcsec. The line-of-sight integration of the observed quantities described above was performed to a depth of 10~Mpc in radius.

To summarize, the observational data used in our analysis includes two-dimensional images of the SZ signal, X-ray SB, and X-ray temperature. Then we used our triaxial model to generate analogous images based on the model parameters delineated in Table~\ref{tab:params}. The observed and model-generated images can then be directly compared to facilitate our fitting process, and the method employed for this fitting procedure is elaborated upon in the following section.

\subsection{Fitting formalism}

The $\chi^2$ statistic is used to define the likelihood of the model. We used \texttt{emcee} \citep{2013PASP..125..306F}, a Python-based affine-invariant ensemble Markov chain Monte Carlo \citep[MCMC;][]{2010CAMCS...5...65G} package, for the model fitting process. By performing MCMC sampling \citep{2018ApJS..236...11H}, we determined the posterior distribution of the parameters that describe the triaxial model. When conducting a model fit with the data, we occasionally needed to adjust the scale parameter of the stretch move within the affine-invariant ensemble sampling algorithm implemented in the package to enhance performance \citep{2015arXiv150902230H}. 

We define the $\chi^2$ functions for our analysis below; they are based on two-dimensional maps of the SZ and X-ray data rather than the original one-dimensional radial profiles used in the CLUMP-3D method presented in \cite{2017MNRAS.467.3801S}. The $\chi^2$ function for the two-dimensional SZ map is
\begin{equation}
    \chi^2_{\text{\tiny SZ}} = \sum_{i,j=1}^{N_{\text{\tiny y}}} \Bigl[y_{i} - \hat{y}_{i} \Bigr] \Bigl( C^{-1}_{\text{\tiny SZ}} \Bigr)_{ij} \Bigl[y_{j} - \hat{y}_{j} \Bigr],
\end{equation}
where $\hat{y}_{i}$ is the model Compton-$y$ within a pixel, and ${y}_{i}$ is the observed value. To deal with the correlated noise in the SZ data, we used the inverse of the uncertainty covariance matrix ($C^{-1}_{\text{\tiny SZ}}$). Similarly, the $\chi^2$ function for the X-ray temperature map becomes
\begin{equation}
    \chi^2_{\text{\tiny T}} = \sum_{i=1}^{N_{\text{\tiny T}}} \left(\frac{T_{\text{\tiny sp}, i} - \hat{T}_{\text{\tiny sp}, i}}{\delta T_{\text{\tiny sp},i}} \right)^2,
\end{equation}
where $\hat{T}_{\text{\tiny sp}, i}$ is the model spectroscopic temperature within a pixel, and ${T}_{\text{\tiny sp}, i}$ is the observed value with uncertainty $\delta T_{\text{\tiny sp},i}$.

For the X-ray SB, we employed a dual approach. We used a two-dimensional model fit within the circular region that encloses 80\% of the emission and one-dimensional analysis for the outside region where the background and the source emission is comparable and signal-to-noise ratio is relatively low. In the exterior region, we computed azimuthal medians in annular bins to mitigate biases in measuring X-ray SB caused by gas clumping, as suggested by \cite{2015MNRAS.447.2198E}. While our current analysis solely uses the two-dimensional map of X-ray temperature, in future work we intend to implement an approach that is fully consistent with our treatment of the X-ray SB to also mitigate local deviations from homogeneity in the X-ray temperature data \citep{2023arXiv231102176L}. Then, the combined likelihood becomes
\begin{equation}
    \chi^2_{\text{\tiny SB}} = \chi^2_{\text{\tiny SB,1D}} + \chi^2_{\text{\tiny SB,2D}}
\end{equation}
where
\begin{equation}
    \chi^2_{\text{\tiny SB,1D}} = \sum_{i=1}^{N_{\text{\tiny SB,1D}}} \left(\frac{S_{\text{\tiny X,1D}, i} - \hat{S}_{\text{\tiny X,1D}, i}}{\delta S_{X,1D,i}} \right)^2,
\end{equation}
and
\begin{equation}
    \chi^2_{\text{\tiny SB,2D}} = \sum_{i=1}^{N_{\text{\tiny SB,2D}}} \left(\frac{S_{\text{\tiny X,2D}, i} - \hat{S}_{\text{\tiny X,2D}, i}}{\delta S_{\text{\tiny X,2D},i}} \right)^2.
\end{equation}
Here $\hat{S}_{\text{\tiny X}, i}$ is the model SB, and $S_{\text{\tiny X}, i}$ and $\delta_{S,i}$ are obtained from the observational data. We employed SB measurements and the corresponding error for our two-dimensional analysis here, assuming Gaussian statistics. This should be a valid assumption, as we define regions with sufficiently large photon counts (i.e., $\ge 20$). However, the formally correct approach is to use the Cash statistic, which accounts for Poisson fluctuations in the photon counts \citep{1979ApJ...228..939C}. Fits using the Cash statistic for photon counting in the low count regime will be explored in future works.

Finally, the total $\chi^2$ statistic becomes
\begin{equation}
    \label{eq:chisq}
    \chi^2_{\text{\tiny X+SZ}} = \chi^2_{\text{\tiny SZ}} + \chi^2_{\text{\tiny T}} + \chi^2_{\text{\tiny SB}},
\end{equation}
and the MCMC is used to sample $\chi^2_{\text{\tiny X+SZ}}$ within the parameter space near the best fit. 
\subsection{Parameter estimation with mock data}\label{sec:mocktest}

To validate the accuracy of our model fitting algorithm, we conducted a full analysis using mock observations of a galaxy cluster described by our model from known input parameter values. Using the model parameters outlined in Table~\ref{tab:params}, we generate model SZ, X-ray SB and temperature maps, incorporating the instrument PSF response. For this test, we generated a wide range of mock datasets by varying the inclination angle ($\cos \theta$) and minor-to-major axial ratio of the ICM distribution ($q_{\text{\tiny ICM,1}}$), using the reference model as a basis. The reference mock cluster has the following characteristics: $z=0.18$, $M_{500}= 8.8 \times 10^{14} \textrm{M}_\odot$, $R_{500}= 7.4$\arcmin. Additionally, we used the following values for the geometric configuration and electron density and pressure parameters, with $(q_{\text{\tiny ICM,1}}, q_{\text{\tiny ICM,2}}, \cos \theta, \varphi, \psi) = (0.6, 0.75, 0.8, -25, 60)$, $(n_0, \zeta_\text{\tiny c}, \zeta_t, \beta_e, \eta_e, \gamma_e) = (0.002, 175, 1.5, 0.6, 0.3, 1.8)$, $P_0, \alpha_p = (10.0, 1.0)$. In this case, $e_\parallel$ = 1.02.

\begin{figure}[t!]
    \centering
    \includegraphics[width=0.33\textwidth]{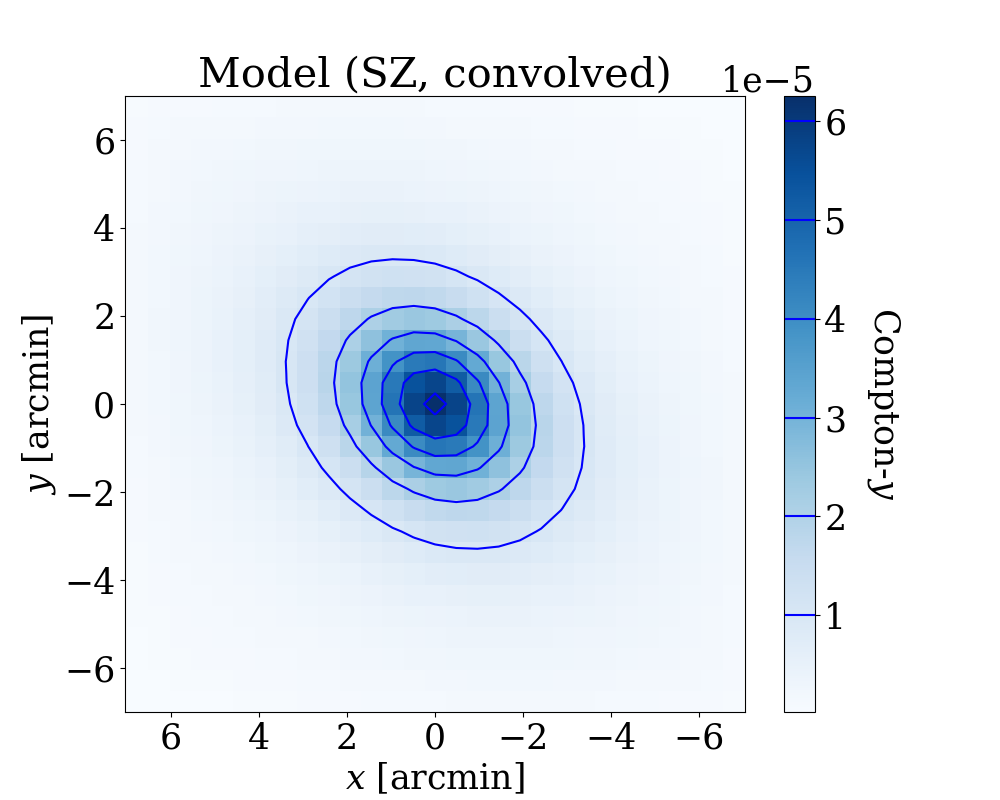}
    \includegraphics[width=0.33\textwidth]{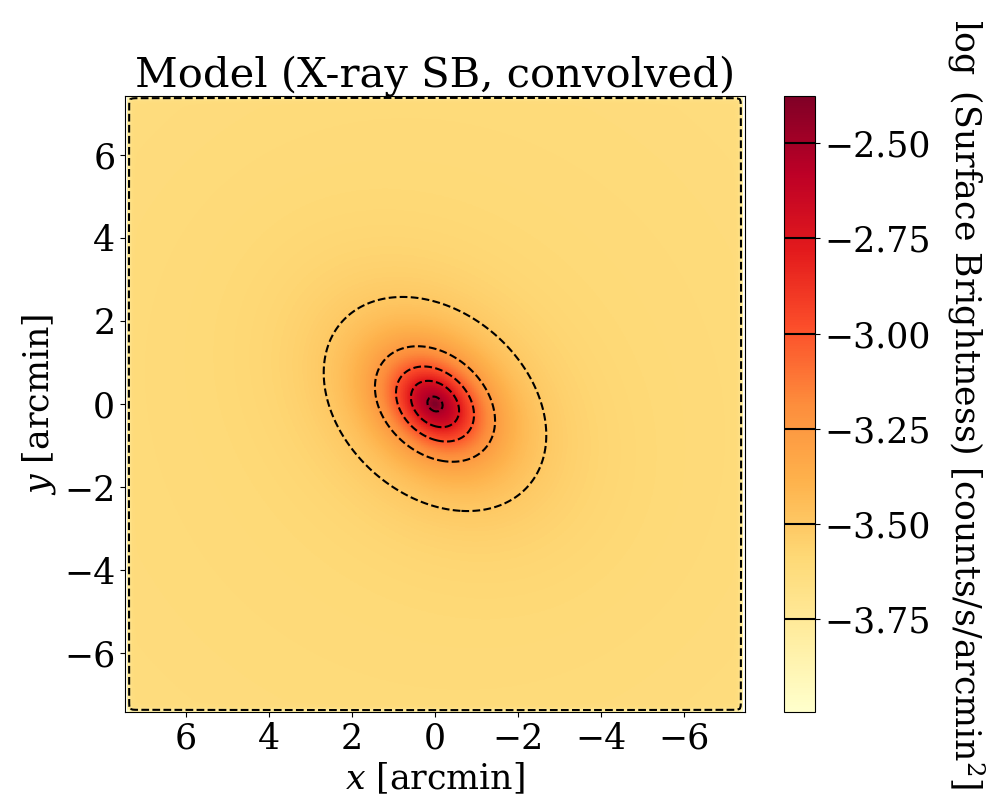}
    \includegraphics[width=0.33\textwidth]{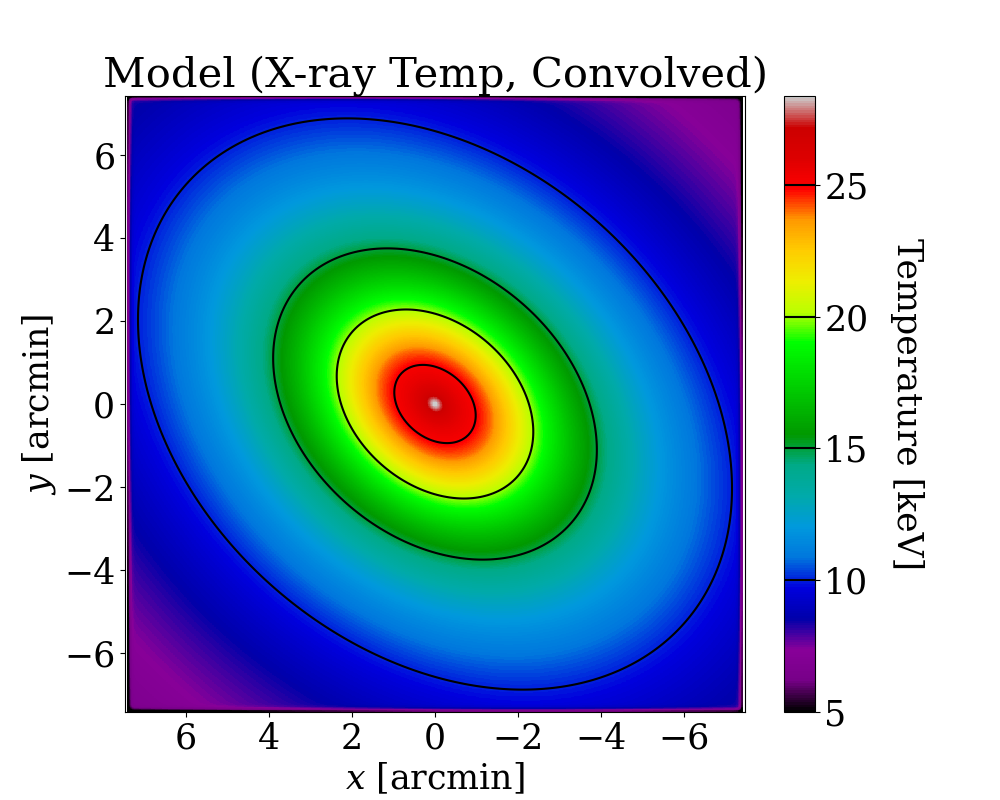}
    \caption{Projected PSF-convolved SZ model map ({\it top}) and PSF convolved X-ray SB ({\it middle}) and temperature ({\it bottom}) maps. The contours for the models are overlaid to improve the visual representation of the maps. The simulated PSFs used for the SZ and X-ray maps correspond to 1.6\arcmin\ and 10\arcsec\ FWHM, respectively. To ensure accurate PSF convolution, we binned the pixels in the maps such that the FWHM of each instrument's PSF is covered by at least three pixels.
    }
    \label{fig:paramtest_model}
\end{figure}

\begin{figure*}[t!]
    \centering
    \includegraphics[width=0.9\textwidth]{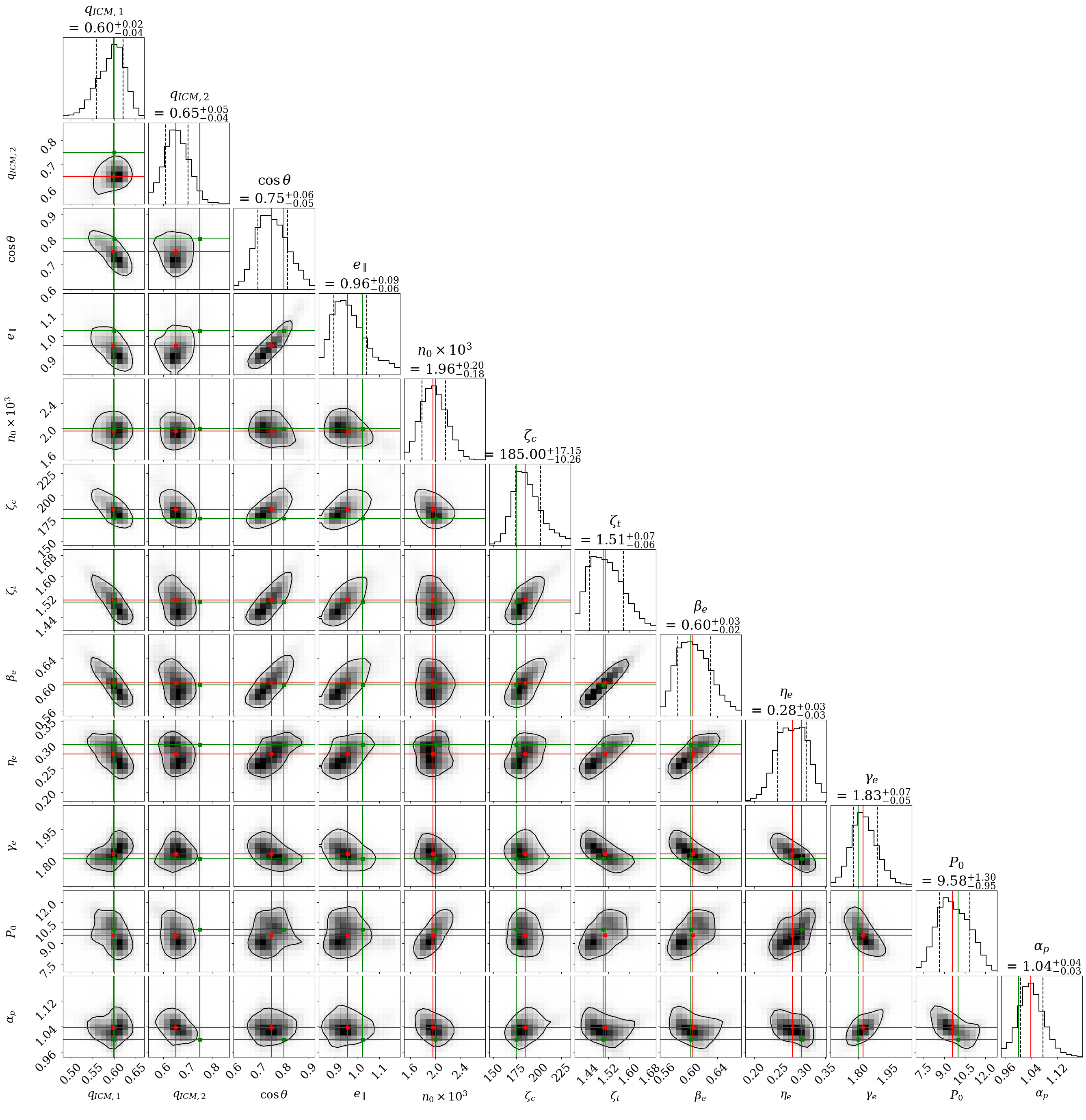}
    \caption{Posterior distributions estimated from our MCMC for a mock observation generated from a smooth model. The vertical green lines in each plot indicate the input parameters used to generate the mock observation maps, while the vertical red lines represent the median value from the accepted MCMC samples. The values displayed above each histogram show the median of the distribution, along with the 1$\sigma$ (68.3\%) credible region, which is indicated by dashed vertical lines in every plot. Additionally, the solid black line in the two-dimensional distributions encloses the 68\% credible region for the parameter pairs. As highlighted by \cite{2006ApJ...640..691V} and \cite{2007ApJ...668....1N}, there is a correlation between the model parameters related to the ICM radial profiles, and individual parameter values exhibit a degeneracy. However, our objective is to ascertain smooth analytic functions that accurately represent the electron density and pressure profiles, thereby providing a comprehensive description of the ICM thermodynamics.
    }
    \label{fig:paramtest_fit}
\end{figure*}

The model maps generated with the input parameters of the reference model are presented in Fig.~\ref{fig:paramtest_model}. For each pixel based on its coordinates within the observed map, we calculated the observables projected onto the two-dimensional sky plane (Sect.~\ref{subsec:observables}). Then, instrumental effects, including the PSF response, are applied. As we discuss in the next section, our baseline analysis of the observed data uses the combined \Planck\ and ACT SZ effect map \citep{2020JCAP...12..047A}, and we assumed a PSF with a FWHM of 1.6\arcmin. To ensure adequate angular sampling of the PSF, we required a maximum pixel size equal to the FWHM divided by 3.

In addition, we incorporated noise into each mock observation. Using the error maps for the observed data, we randomly sampled Gaussian noise distributions for the SZ, X-ray SB, and X-ray temperature maps, respectively. 
Figure~\ref{fig:paramtest_fit} shows the posterior distribution of the parameters from our fit to this mock observation. The posterior distributions indicate that we can accurately recover most of the varied parameter values within the expected deviations due to noise fluctuations. Thus, our fitting methodology is able to reliably determine the underlying shape and thermodynamics of the observed mock galaxy cluster.

The use of both SZ and X-ray data in our analysis allows us to measure the three-dimensional geometry of the ICM distribution by constraining the elongation parameter (Sect.~\ref{subsec:geom}), since the two observational probes redundantly measure the thermodynamic properties of the gas along the line of sight. However, it should be noted that there may be degeneracies in determining cluster shape through this multi-probe approach depending on the relative orientation of the geometry, especially in inferring the geometric parameters of the three-dimensional structure, as discussed in \cite{2007MNRAS.380.1207S}. These degeneracies can cause bias in the recovered shape parameters along with multi-modality in the posterior distributions. 

\begin{figure}[t!]
    \centering
    \includegraphics[width=0.45\textwidth]{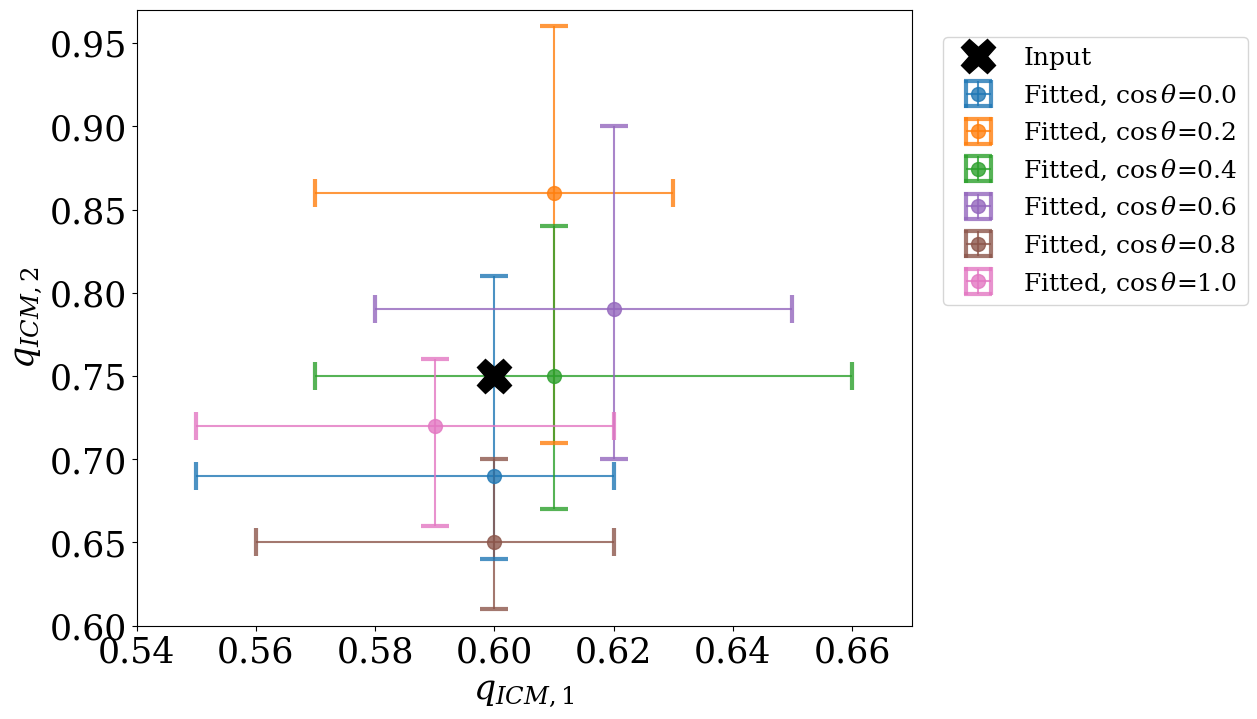}
    \includegraphics[width=0.45\textwidth]{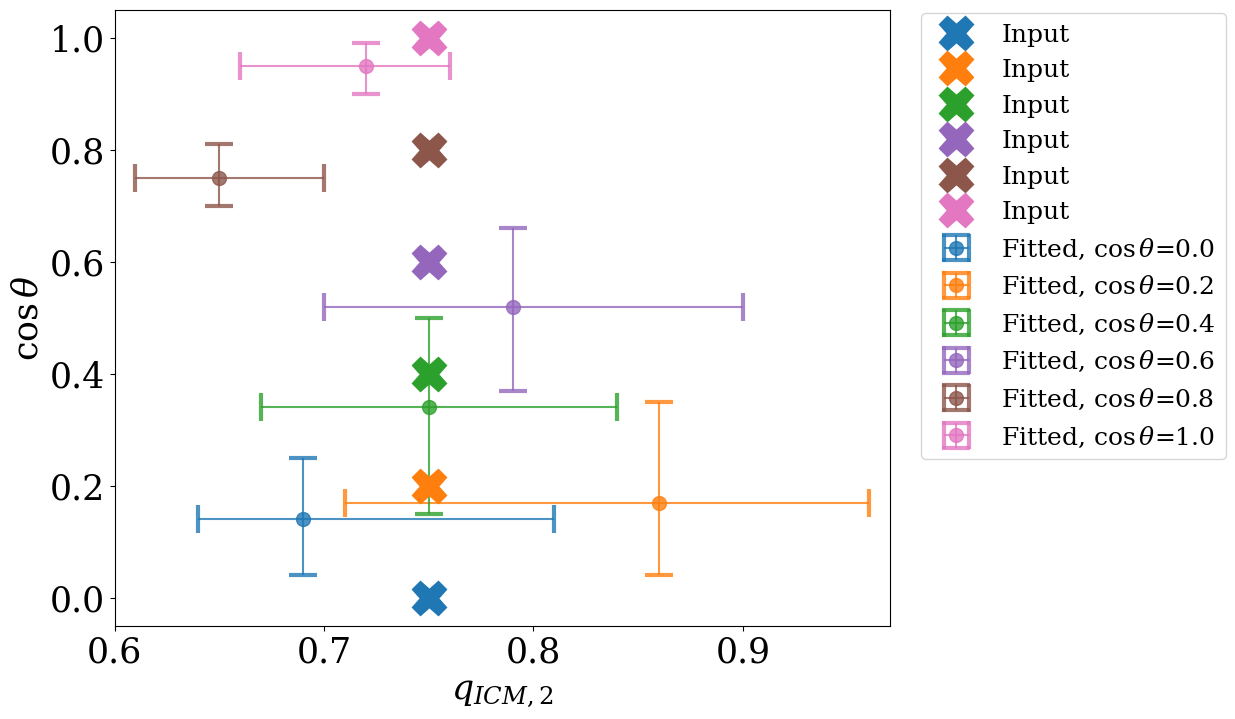}
    \caption{Summary of the model parameters inferred from the mock data, as detailed in Table~\ref{tab:paramtest}. The summary includes comparisons of $q_{\text{\tiny ICM,1}}$ to $q_{\text{\tiny ICM,12}}$ ({\it top}) and $q_{\text{\tiny ICM,2}}$ to $\cos \theta$ ({\it bottom}), along with their estimated uncertainties. The crosses in the figures indicate the input model parameters.}
    \label{fig:paramtest_summary}
\end{figure}

The model fitting technique has been validated using the reference mock dataset, as shown in Fig.~\ref{fig:paramtest_fit}. Additionally, we present the results of parameter estimation across a broader range of parameters. This analysis particularly concentrates on geometric parameters that define the cluster's shape and its projected appearance in the sky. In Table~\ref{tab:paramtest} and Fig.~\ref{fig:paramtest_summary}, we provide constraints for the geometrical parameters of a triaxial ellipsoid ($q_{\text{\tiny ICM,1}}$, $q_{\text{\tiny ICM,2}}$, and $\cos \theta$, $e_\parallel$). These constraints are derived from various model fits, each with different values of $q_{\text{\tiny ICM,1}}$ and $\cos \theta$, while keeping the other parameters constant. We used the model parameters from the previous fit (Fig.~\ref{fig:paramtest_fit}) as a reference for these analyses.

\begin{table*}[!t]
\caption{\label{tab:paramtest} Model parameters and their values estimated using mock data.}
    \centering
    \begin{tabular}{cccccccc}
    \hline\hline
    \multicolumn{8}{c}{Model Parameters (input/triaxial fit results)} \\
    \hline
    \multicolumn{2}{c}{$q_{\text{\tiny ICM,1}}$} & \multicolumn{2}{c}{$q_{\text{\tiny ICM,2}}$} & \multicolumn{2}{c}{$\cos \theta$} & \multicolumn{2}{c}{$e_\parallel$} \\
    \hline
    Input & Fitted & Input & Fitted & Input & Fitted & Input & Fitted \\
    \hline
    0.60 & $0.60^{+0.02}_{-0.05}$ & 0.75 & $0.69^{+0.12}_{-0.05}$ & 0.0 & $0.14^{+0.11}_{-0.10}$ & 0.71 & $0.66^{+0.14}_{-0.07}$ \\
    0.60 & $0.61^{+0.02}_{-0.04}$ & 0.75 & $0.86^{+0.10}_{-0.15}$ & 0.2 & $0.17^{+0.18}_{-0.13}$ & 0.72 & $0.83^{+0.11}_{-0.13}$ \\
    0.60 & $0.61^{+0.04}_{-0.04}$ & 0.75 & $0.75^{+0.09}_{-0.08}$ & 0.4 & $0.34^{+0.16}_{-0.19}$ & 0.77 & $0.75^{+0.09}_{-0.06}$ \\
    0.60 & $0.62^{+0.03}_{-0.04}$ & 0.75 & $0.79^{+0.11}_{-0.09}$ & 0.6 & $0.52^{+0.14}_{-0.15}$ & 0.86 & $0.85^{+0.08}_{-0.10}$ \\
    0.60 & $0.60^{+0.02}_{-0.04}$ & 0.75 & $0.65^{+0.05}_{-0.04}$ & 0.8 & $0.75^{+0.06}_{-0.05}$ & 1.02 & $0.96^{+0.09}_{-0.06}$ \\
    0.60 & $0.59^{+0.03}_{-0.04}$ & 0.75 & $0.72^{+0.04}_{-0.06}$ & 1.0 & $0.95^{+0.04}_{-0.05}$ & 1.33 & $1.28^{+0.07}_{-0.09}$ \\
    \hline
    0.15 & $0.13^{+0.03}_{-0.02}$ & 0.75 & $0.80^{+0.09}_{-0.10}$ & 0.8 & $0.75^{+0.05}_{-0.05}$ & 0.59 & $0.45^{+0.13}_{-0.07}$ \\
    0.30 & $0.29^{+0.03}_{-0.02}$ & 0.75 & $0.71^{+0.07}_{-0.09}$ & 0.8 & $0.81^{+0.06}_{-0.08}$ & 0.87 & $0.87^{+0.12}_{-0.14}$ \\
    0.45 & $0.44^{+0.03}_{-0.04}$ & 0.75 & $0.68^{+0.09}_{-0.08}$ & 0.8 & $0.80^{+0.08}_{-0.09}$ & 0.98 & $0.96^{+0.11}_{-0.11}$ \\
    0.60 & $0.60^{+0.02}_{-0.04}$ & 0.75 & $0.65^{+0.05}_{-0.04}$ & 0.8 & $0.75^{+0.06}_{-0.05}$ & 1.02 & $0.96^{+0.09}_{-0.06}$ \\
    \hline
    \end{tabular}
\end{table*}

In general, the input model parameters are successfully recovered within the test parameter ranges, although the uncertainties in the inclination angle increase as the major axis of the triaxial ellipsoid approaches a perpendicular orientation relative to the observer's line of sight. Most of the input values fall within the 68\% confidence interval obtained from our fits, while the remaining values are included within the 95\% confidence interval. Considering the expected fluctuations caused by the random noise added to the input maps, this suggests that any potential bias in our fitting method is minimal compared to the measurement noise. In a subsequent paper, we will perform a more detailed exploration spanning the range of masses, redshifts, and data quality in the CHEX-MATE sample. In addition, we will fit mock observations of simulated clusters to determine the impact of deviations from triaxial symmetry due to, for example, sub-structures or mergers.



\section{Application to CHEX-MATE data}\label{sec:application_chexmate}

In this section, we introduce the X-ray and SZ data collected from our program. We applied the triaxial analysis technique to analyze a CHEX-MATE galaxy cluster PSZ2 G313.33+61.13 (Abell 1689), and the cluster serves as an illustrative example to demonstrate the method.

\subsection{Data}

Table~\ref{tab:data} summarizes the SZ and X-ray data from CHEX-MATE available for our multiwavelength analysis of the ICM distribution. The foundation of our analysis is the 3~Msec \xmm\ observing program CHEX-MATE \citep{2021A&A...650A.104C}, from which we obtained two-dimensional X-ray SB and temperature maps produced using the Voronoi tessellation method \citep{2003MNRAS.342..345C, 2006MNRAS.368..497D}. The details of the image production are reported in \cite{2023AnA...674A.179B} and \cite{2023arXiv231102176L}, and here we report briefly the main analysis steps.

\begin{table*}[!t]
\caption{\label{tab:data} X-ray and SZ observation data and instruments used for the analysis of PSZ2 G313.33+61.13.}
    \centering
    \begin{tabular}{cccl}
    \hline\hline
    Wavelength & Type & Instrument & Reference \\
    \hline
    X-ray & Surface brightness (SB) & \xmm\ & \cite{2023AnA...674A.179B} \\
    X-ray & Temperature & \xmm\ & \cite{2023arXiv231102176L} \\
    mm-wave & SZ $y$-map & \Planck\ \tablefootmark{a} & \cite{2016AA...594A..22P} \\
    mm-wave & SZ $y$-map & ACT (ACTPol)\tablefootmark{b} & \cite{2020PhRvD.102b3534M} \\
    \hline
    \end{tabular}
    \tablefoot{The publicly released datasets of the surveys are available in the links below.\\
    \tablefoottext{a}{\url{https://heasarc.gsfc.nasa.gov/W3Browse/all/plancksz2.html}}\\
    \tablefoottext{b}{\url{https://github.com/ACTCollaboration/DR4_DR5_Notebooks}}\\
    }
\end{table*}

The \xmm\ observations of the clusters were obtained using the EPIC instrument \citep{2001A&A...365L..18S, 2001A&A...365L..27T}. To create the X-ray SB map, photon-count images in the [0.7-1.2] keV range were extracted from the data acquired using the MOS1, MOS2, and pn cameras on the instrument. The energy band was selected to optimize the contrast between the emission from the source and the background \citep{2010A&A...524A..68E}. The images from all three cameras were combined to maximize the statistical significance while accounting for the energy-band responses. Additionally, the X-ray maps are instrumental-background subtracted and corrected for the exposure. Point sources are removed from the analysis \citep{2019A&A...621A..41G} by masking them with circular regions that appear as empty circular regions in the X-ray maps in Fig.~\ref{fig:psz2g313-input}. Furthermore, they are spatially binned to have at least 20 counts per bin using the Voronoi technique. X-ray temperature maps \citep{2023arXiv231102176L} were prepared in a similar manner for the data obtained in the [0.3-7] keV band, with background modeling \citep{2019MNRAS.483..540L} and spectral fitting performed. The fitting procedure to ascertain the temperature was done utilizing XSPEC \citep{1996ASPC..101...17A}, which was employed to minimize the modified Cash statistics \citep{1979ApJ...228..939C} with the assumption of \cite{2009ARA&A..47..481A} metallicity. Subsequently, Voronoi-binned maps were generated to achieve a high signal-to-noise ratio ($\sim$30) for each cell.

\Planck\ SZ maps are available for all of the CHEX-MATE galaxy clusters by definition \citep{2013A&A...550A.131P, 2021A&A...651A..73P}. From these data we generated a custom $y$-map using the Modified Internal Linear Component Algorithm \citep[MILCA;][]{2013A&A...558A.118H} with an improved angular resolution of 7\arcmin\ FWHM compared to the one publicly released by \Planck\ with an angular resolution of 10\arcmin\ FWHM \citep{2016AA...594A..22P}. Also, ground-based SZ observations from cosmic microwave background surveys, including the ACT and the South Pole Telescope \citep[SPT;][]{2022ApJS..258...36B}\footnote{\url{https://pole.uchicago.edu/public/data/sptsz_ymap/}}, as well as the Caltech Submillimeter Observatory (CSO) Bolocam galaxy cluster archive\footnote{\url{https://irsa.ipac.caltech.edu/data/Planck/release_2/ancillary-data/bolocam/bolocam.html}} \citep{2013ApJ...768..177S}, provide higher angular resolution data for a subset of CHEX-MATE clusters. Some of these ground-based data are currently publicly accessible, while others are slated for future release.

In this demonstration paper, we make use of the ACT SZ component-separated maps. The recent data release 4 (DR4) from the ACT provides component-separated maps, one of which is the SZ \citep{2020JCAP...12..047A, 2020PhRvD.102b3534M}. These maps were generated by analyzing data from a 2,100 square degree area of the sky, captured using the ACTPol receiver \citep{2016JLTP..184..772H} at 98 and 150 GHz. These data offer more than four times finer angular resolution compared to the \Planck\ map. Then, the maps were jointly analyzed and combined with \Planck\ data. Rather than using the noise estimate provided with these data, which is quantified as a two-dimensional power spectral density, we instead followed an approach based on the recent analysis of similar joint ACT and \Planck\ maps in \cite{2021A&A...651A..73P}. Specifically, we randomly sampled 10,000 maps, ensuring that their size aligns with that of the input SZ data, in the corresponding ACT region (for instance, the region designated as ``BN'' for the cluster Abell 1689 analyzed in the next section). Then, we computed the covariance using these maps to estimate the noise covariance matrix. The resulting noise rms for the $y$-map is approximately $\sim 9 \times 10^{-6}$ per 0.5\arcmin\ square pixel, and the diagonal elements of the noise covariance matrix are shown along with the $y$-map in Fig.~\ref{fig:psz2g313-input}.





\subsection{PSZ2 G313.33+61.13 (Abell 1689)}

\begin{figure*}[t!]
    \centering
    \includegraphics[width=.33\textwidth]{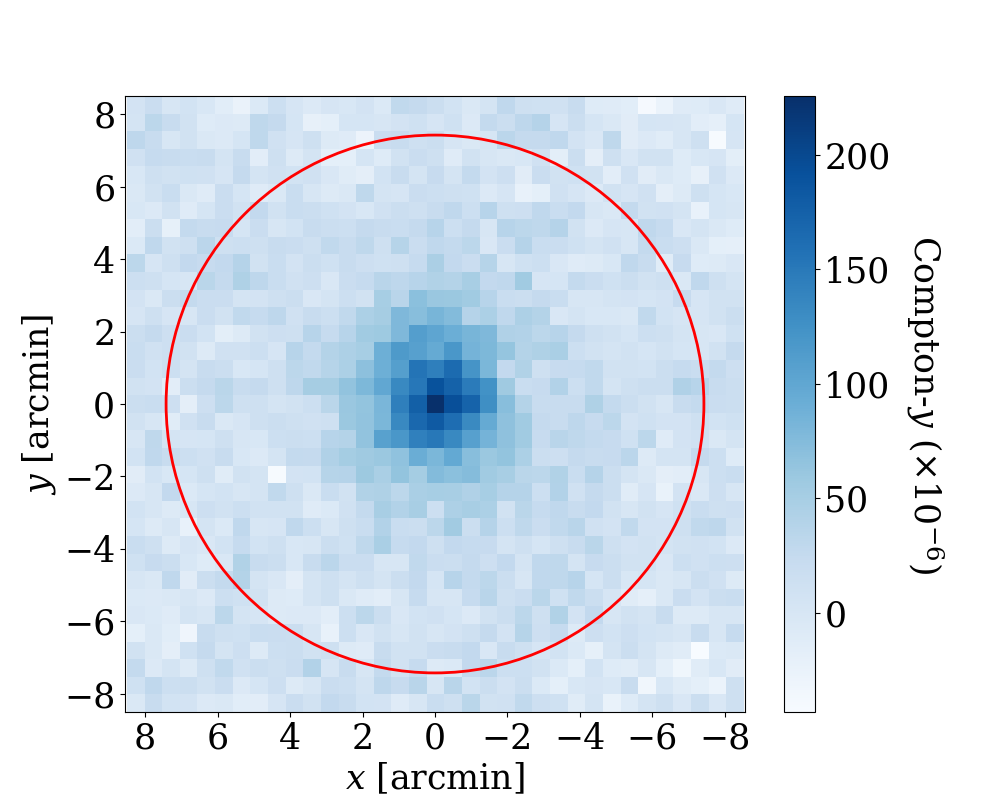}\hfill
    \includegraphics[width=.33\textwidth]{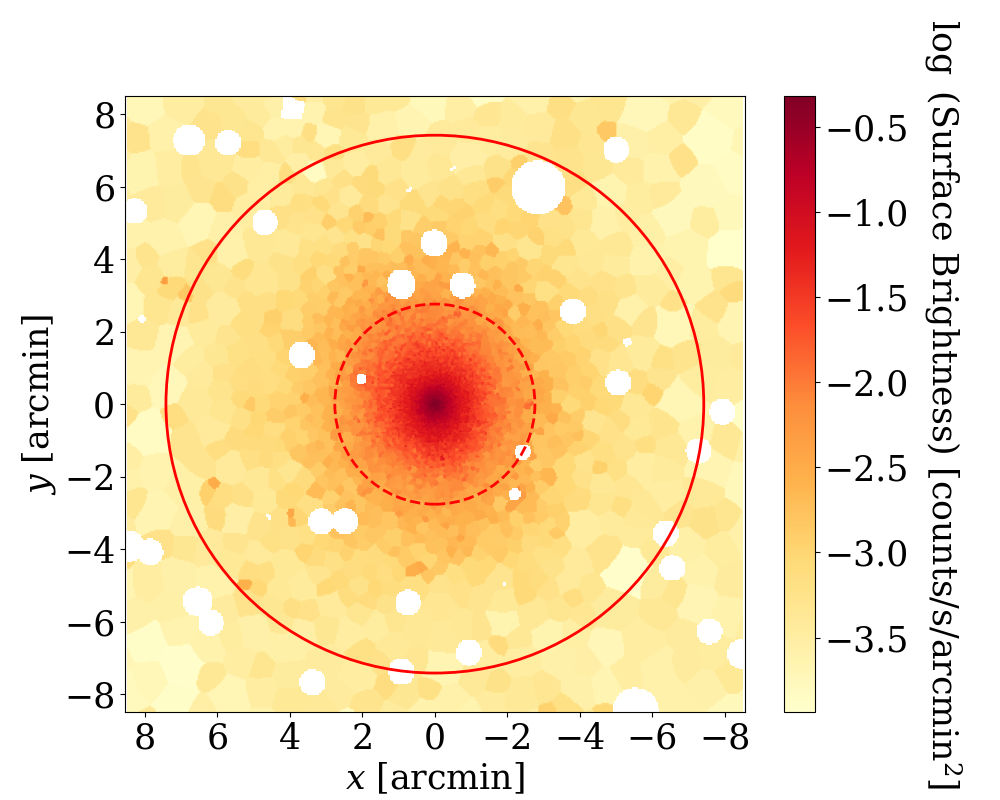}\hfill
    \includegraphics[width=.33\textwidth]{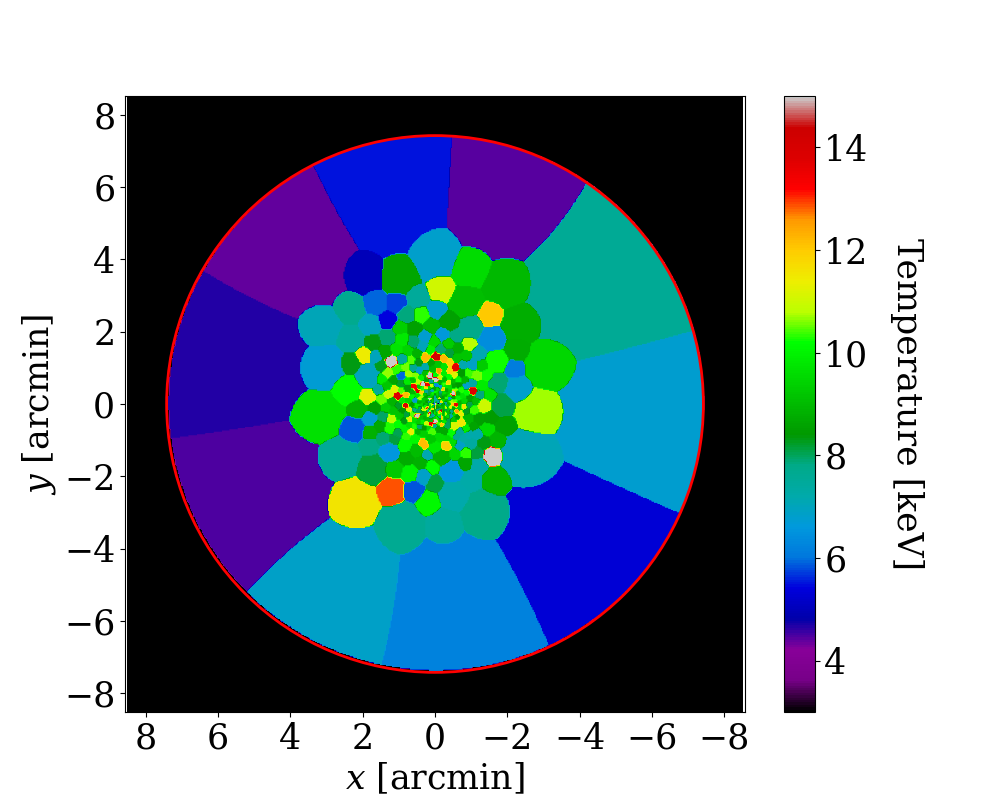}\\
    \includegraphics[width=.33\textwidth]{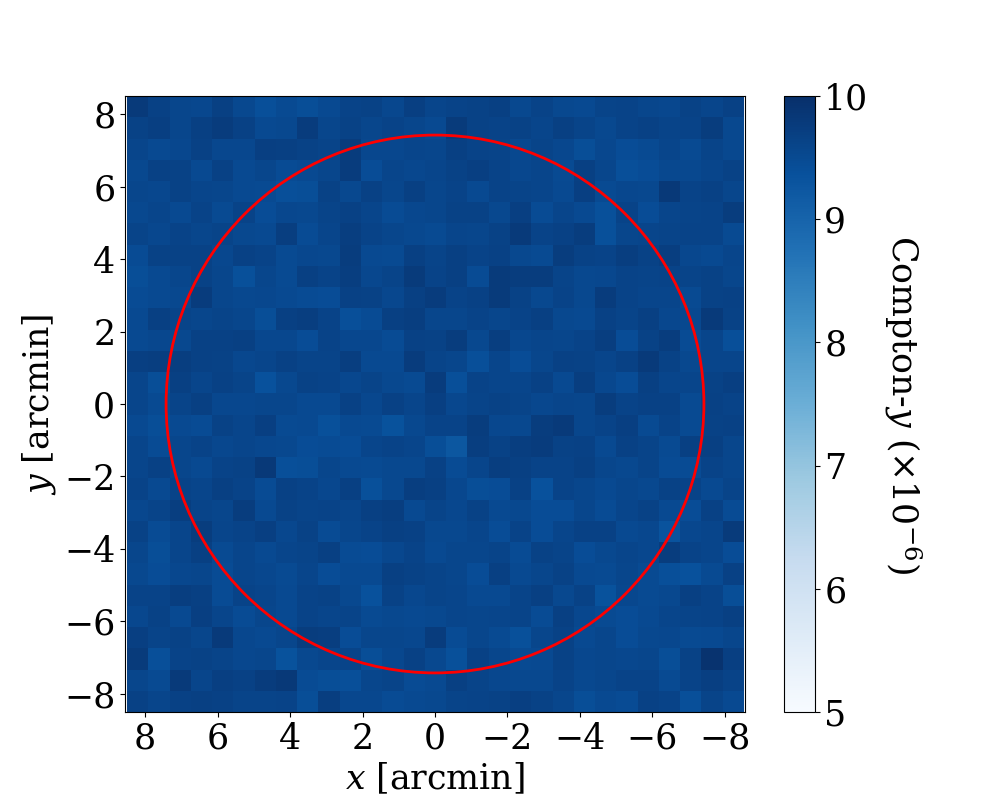}\hfill
    \includegraphics[width=.33\textwidth]{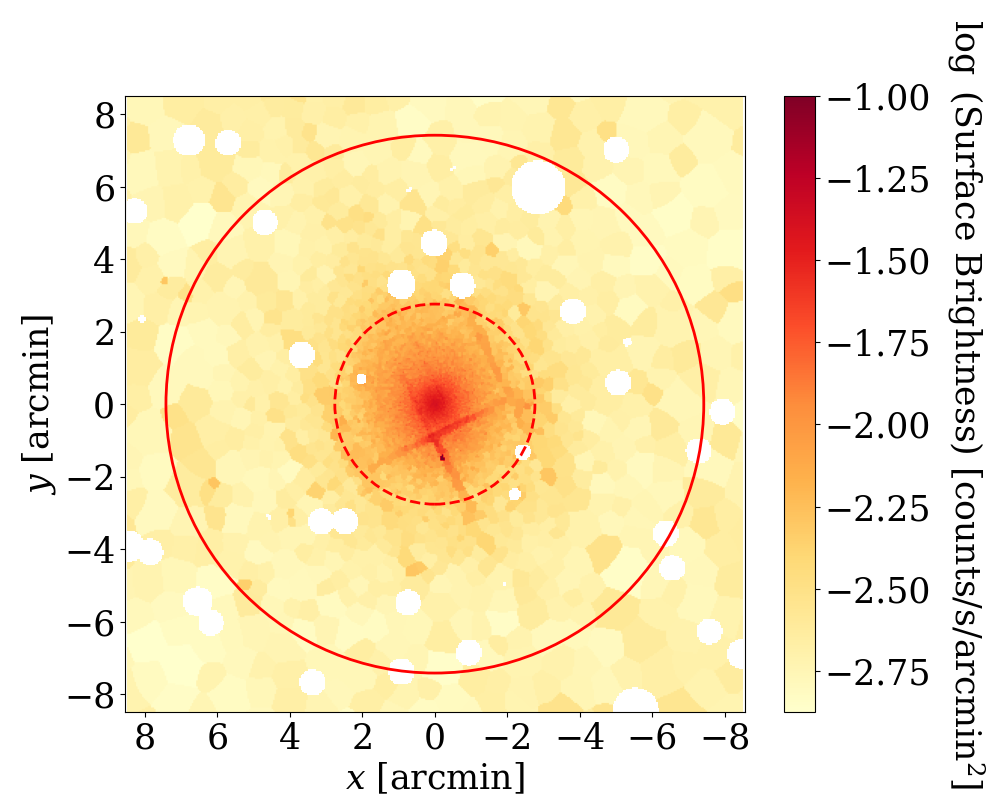}\hfill
    \includegraphics[width=.33\textwidth]{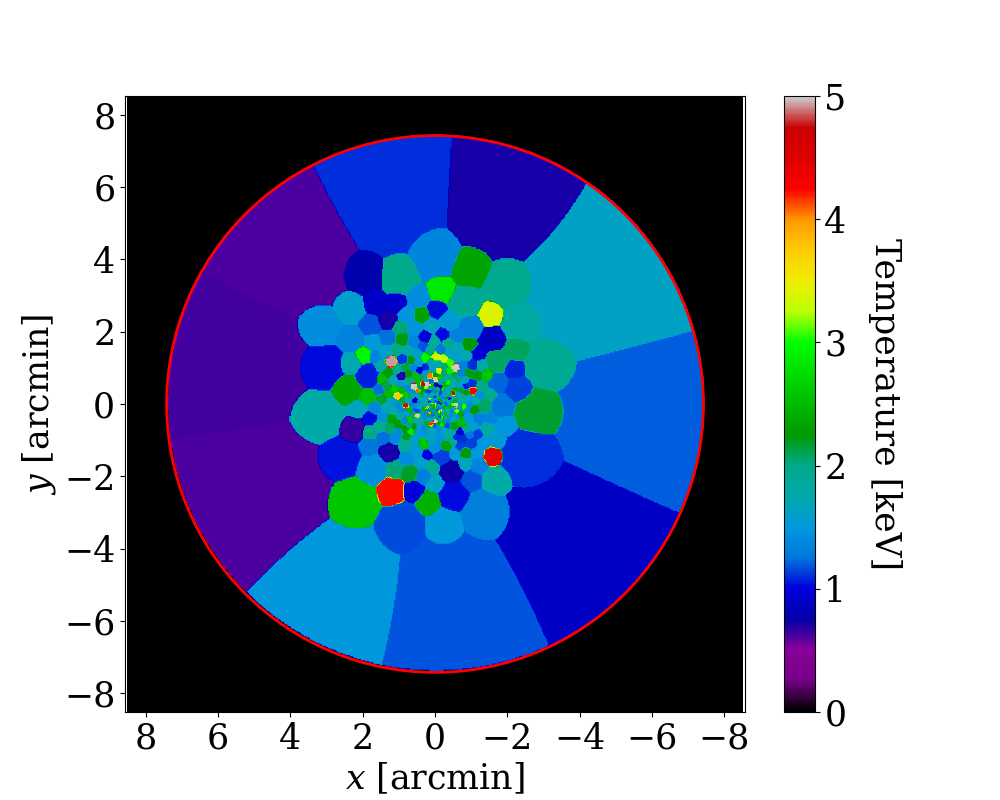}
    \caption{SZ and X-ray maps of the CHEX-MATE galaxy cluster PSZ2 G313.33+61.13 (Abell 1689). The ACT Compton-$y$ map ({\it top left}) and its error map ({\it bottom left}), the X-ray SB map ({\it top middle}) and its error map ({\it bottom middle}), and the X-ray temperature map ({\it top right}) and its error map ({\it bottom right}) are shown. The ACT SZ map is one of the component-separated map products that were produced using the internal linear combination method and combined with data from \Planck\ (i.e., this is a joint map from ACT + \Planck\ data; \citealt{2020PhRvD.102b3534M}). The X-ray maps are the data products from \cite{2023AnA...674A.179B} and \cite{2023arXiv231102176L}. Bright point sources in the X-ray SB maps are indicated by white circles and were masked in the analysis, and the same point-source regions were also removed from the spectral analysis to obtain the temperatures. The regions are excluded from the model fit. The X-ray SB maps were binned using a Voronoi algorithm to ensure an adequate number of photon counts per bin, with smaller bin sizes used in the central region where the count rates are higher. For the temperature maps, Voronoi binning was similarly applied using a fixed signal-to-noise ratio of 30 instead of a fixed number of counts, ensuring a roughly uniform statistical uncertainty per bin. In the X-ray and SZ maps, red circles indicate the two-dimensional map regions included in our analysis. We incorporated a circular region with a radius of $r=R_{500}$ around the galaxy cluster center for the SZ and X-ray data by applying a circular mask to the maps, and the radius is shown as red circles. For this particular cluster, $R_{500}$ is equal to 7.42\arcmin\ ($\sim$1.37~Mpc). In the X-ray SB and its corresponding error maps, dashed red circles represent the region that encompasses 80\% of the emission in the SB map, which is where the two-dimensional (inner region) and one-dimensional (outer region) analyses are separated, and it is located at $r=2.58$\arcmin. We will explore and implement a comparable approach that integrates both two- and one-dimensional analysis techniques for temperature data, as described by \citet{2023arXiv231102176L}, in a forthcoming analysis.}
    \label{fig:psz2g313-input}
\end{figure*}

Using the datasets described above, we demonstrate our fitting method for PSZ2 G313.33+61.13 (Abell 1689), which is a Tier-2 cluster in the CHEX-MATE sample located at $z=0.1832$ with a \Planck\ SZ estimated mass of $M_{500} = 8.77 \times 10^{14} \textrm{M}_\odot$. We note the lensing mass measurement of the cluster is $\sim$70\% higher than the \Planck\ hydrostatic mass estimate \citep[see][]{2015ApJ...806..207U}. We conducted a triaxial fit aligning the model center with the X-ray peak \citep{2023AnA...674A.179B}. For a morphologically regular cluster, like Abell 1689, any deviations or offsets between the SZ and X-ray measurements are expected to have minimal impact on the overall model fit. The \Planck\ + ACT SZ $y$-maps, along with the \xmm\ X-ray SB and temperature maps, are shown in Fig.~\ref{fig:psz2g313-input}.  Maps of the rms noise for each observable are also included, and indicate that the cluster is imaged at high signal to noise. 

In the \xmm\ maps presented, we masked bright point sources and excluded these areas from our X-ray analysis. As outlined in \cite{2019A&A...621A..41G} and \cite{2023AnA...674A.179B}, the X-ray images were prepared to ensure a consistent cosmic X-ray background flux across the entire field of view during the point source subtraction process. Consequently, X-ray sources that fall below the masking threshold are not expected to have a significant impact on the model fit.

For the SZ data, we identified radio and dusty point sources using the NRAO VLA Sky Survey \citep[NVSS;][]{1998AJ....115.1693C} at 1.4 GHz and the {\it Herschel} SPIRE Point Source Catalog. Our search was centered on the region with $r \leq R_{500}$ around Abell 1689 (RA 197.860 deg, Dec -1.336 deg). Within this area, we found five NVSS radio sources, with the brightest emitting 59.6 mJy. To infer their flux densities at 150~GHz, we identified counterparts in the OVRO/BIMA 30~GHz catalog \citep{2007AJ....134..897C} and calculated the spectral index $\alpha$ (flux density $S \propto \nu^\alpha$), following the methodology described in \cite{2013ApJ...764..152S}. For dusty sources, we performed a similar analysis and fitted a gray-body spectrum ($S \propto \nu^\beta B_\nu(\nu, T)$) to 14 sources that were identified in the 500, 350, and 250~$\mu$m bands of the {\it Herschel} catalogs \citep{2013ApJ...778...52S}. These radio and dusty sources' flux densities, when extrapolated to 150~GHz, were all found to be below $\sim$2~mJy, which is the typical rms noise level of the ACT maps \citep{2020PhRvD.102b3534M}. Specifically, the flux densities of the five radio sources are below 0.16 mJy with the mean of 0.06 mJy, and the 14 dusty sources have an average flux density of 1.04~mJy with a standard deviation of 0.3 mJy, with the brightest at 1.63~mJy. Consequently, given that none of the point sources are found to be brighter than the map noise rms, we determined that point source contamination in the SZ map is negligible, allowing us to use the data without additional masking.

This particular cluster was chosen for our demonstration because its triaxial shape has been well studied in the literature \citep{2011MNRAS.416.2567M, 2011MNRAS.416.3187S, 2012MNRAS.419.2646S, 2015ApJ...806..207U}. For example, \cite{2012MNRAS.419.2646S} performed a gas-only analysis using radial profiles of the X-ray and SZ observations from {\em Chandra}, \xmm\ , and {\em WMAP}, along with various ground-based SZ facilities, and constrained the shape and orientation of the cluster's triaxial model with $q_{\text{\tiny ICM,1}} = 0.70 \pm 0.15$, $q_{\text{\tiny ICM,2}} = 0.81 \pm 0.16$, and $\cos \theta = 0.70 \pm 0.29$. A subsequent study by \citet{2015ApJ...806..207U} presented a combined multiwavelength analysis that included lensing data, with the inferred ICM distribution being $q_{\text{\tiny ICM,1}} = 0.60 \pm 0.14$, $q_{\text{\tiny ICM,2}} = 0.70 \pm 0.16$. Their derived value of $\cos \theta$, obtained from the combined lensing and X-ray/SZ analysis, was found to be $0.93 \pm 0.06$. The large $\cos \theta$ suggests that the major axis of the triaxial ellipsoid ($x^{\text{\tiny int}}_3$ in Fig.~\ref{fig:ellipsoid}) is closely aligned with the observer's line of sight.

\begin{table*}[!t]
\caption{\label{tab:psz2g313-summary} Parameters describing the triaxial geometry of PSZ2 G313.33+61.13 (Abell 1689)}
    \centering
    \begin{tabular}{ccccl}
    \hline\hline
    $q_{\text{\tiny ICM,1}}$ & $q_{\text{\tiny ICM,2}}$ & $\cos \theta$ & $e_\parallel$ & Reference \\
    \hline
    $0.70 \pm 0.15$ & $0.81 \pm 0.16$ & $0.70 \pm 0.29$ & $1.68 \pm 0.53$ &  \cite{2012MNRAS.419.2646S} \\
    $0.60 \pm 0.14$ & $0.70 \pm 0.16$ & $0.93 \pm 0.06$ & $1.16 \pm 0.10$ & \cite{2015ApJ...806..207U} \\
    $0.65 \pm 0.02$ & $0.79 \pm 0.02$ & $\geq 0.96$\tablefootmark{a} & $1.24 \pm 0.03$ & This work \\
    \hline
    \end{tabular}
    \tablefoot{\tablefoottext{a}{at 90\% confidence}
    }
\end{table*}


\begin{figure*}[t!]
    \centering
    \includegraphics[width=0.875\textwidth]{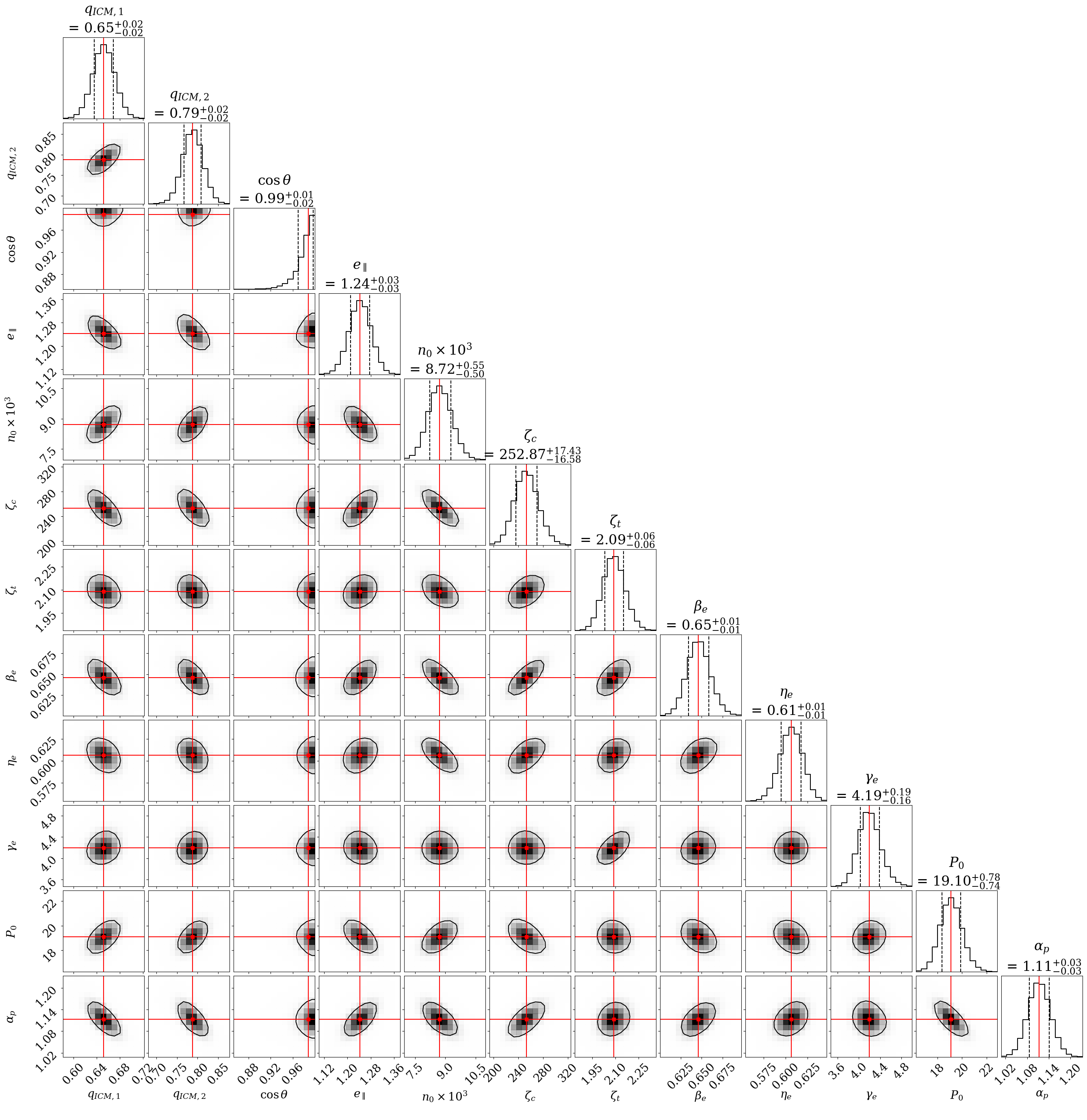}
    \caption{Posterior distribution of the model fit parameters for the galaxy cluster PSZ2 G313.33+61.13 obtained using SZ data from \Planck\ and ACT, as well as X-ray data from \xmm. The  vertical red lines indicate the median value from the accepted MCMC samples, with values displayed along with their 68\% credible regions above each histogram. Instead of the Euler angles $\varphi$ and $\psi$, we present $e_\parallel$, which is a function of five geometric parameters of a triaxial ellipsoid (Eq.~\ref{eq:elongation}).
    }
    \label{fig:psz2g313-corner}
\end{figure*}

\begin{figure*}[t!]
    \centering
    \includegraphics[width=0.33\textwidth]{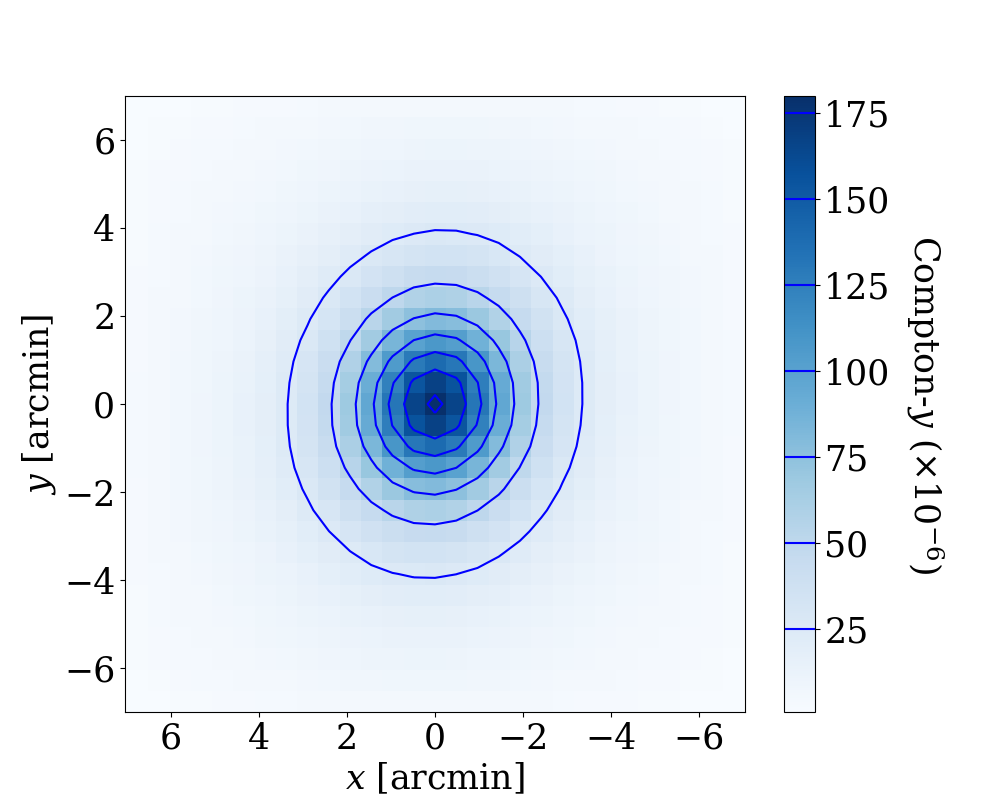}\hfill
    \includegraphics[width=0.33\textwidth]{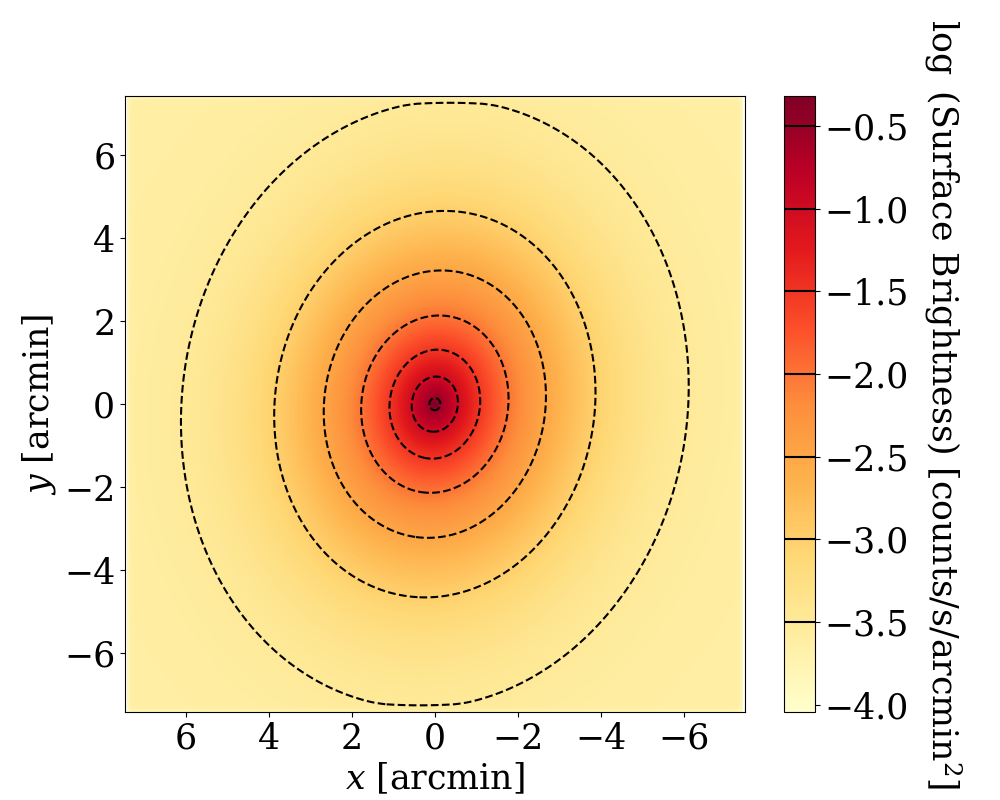}\hfill
    \includegraphics[width=0.33\textwidth]{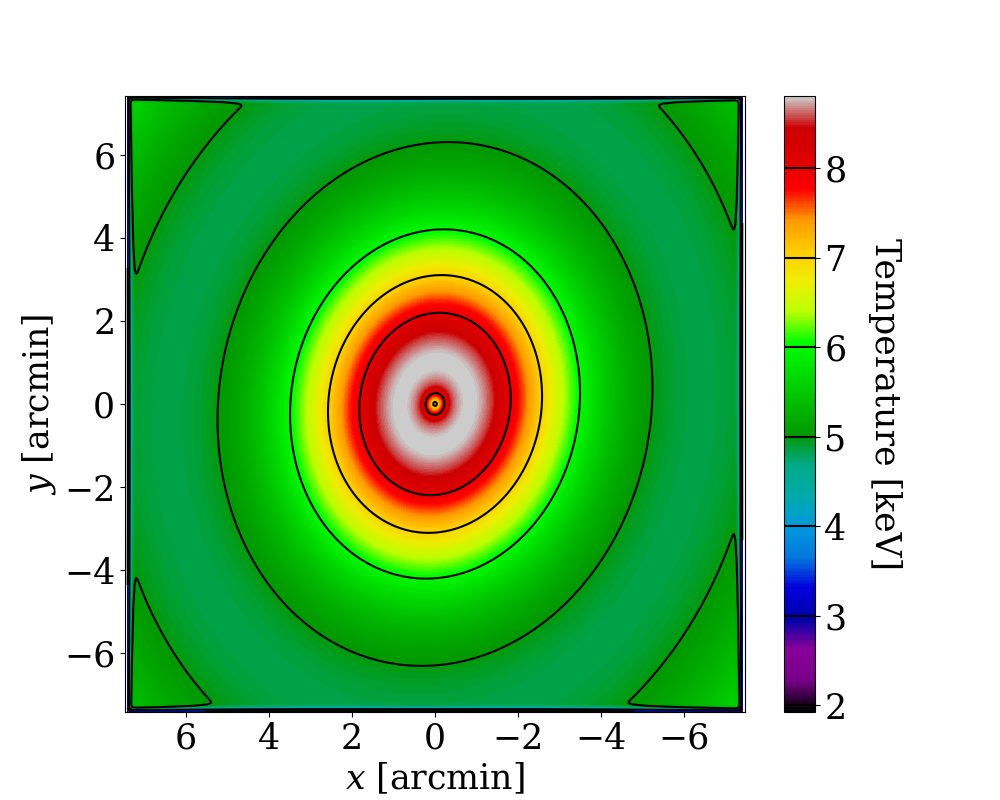}\hfill
    \\
    \includegraphics[width=0.33\textwidth]{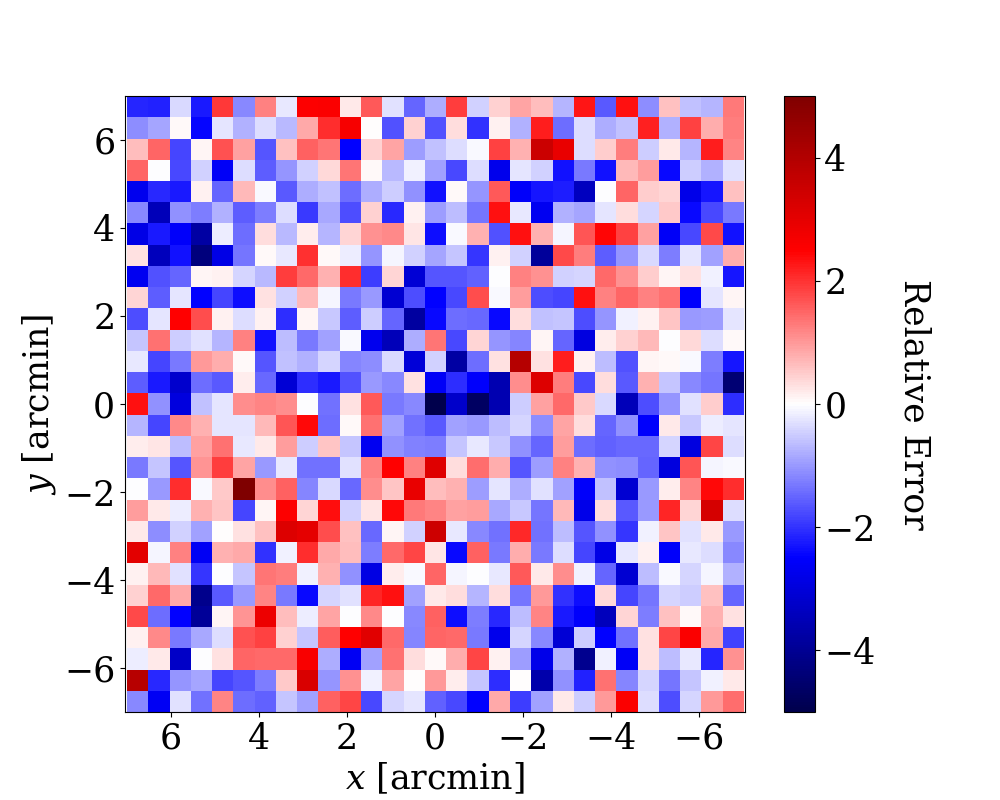}\hfill
    \includegraphics[width=0.33\textwidth]{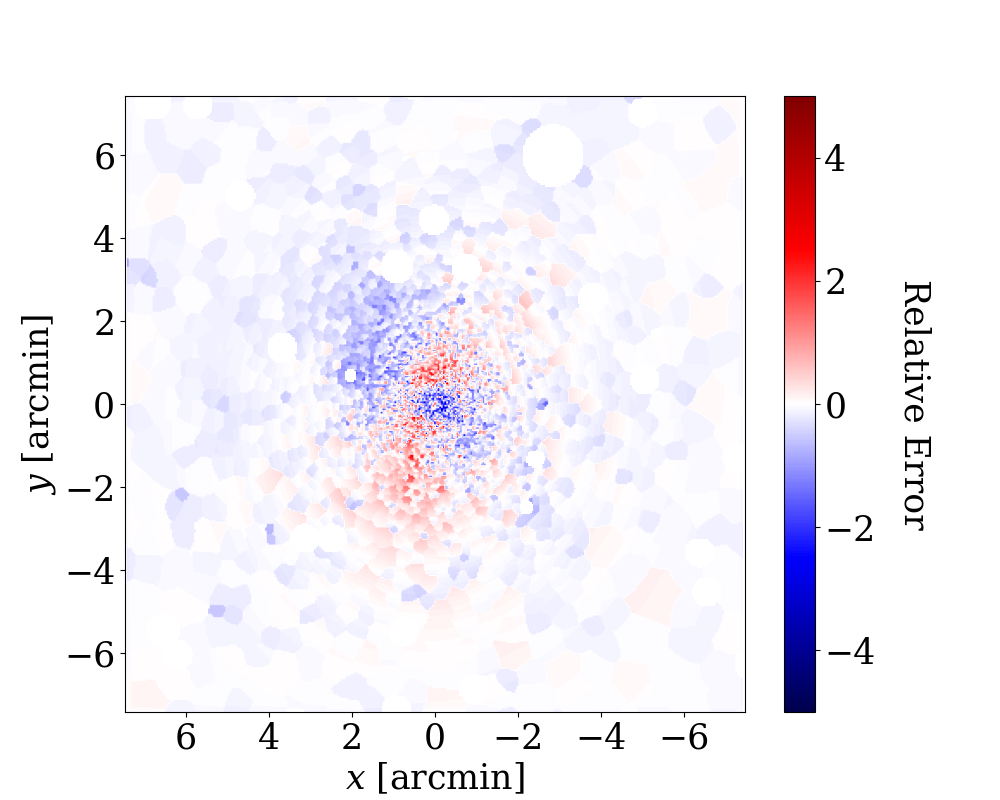}\hfill
    \includegraphics[width=0.33\textwidth]{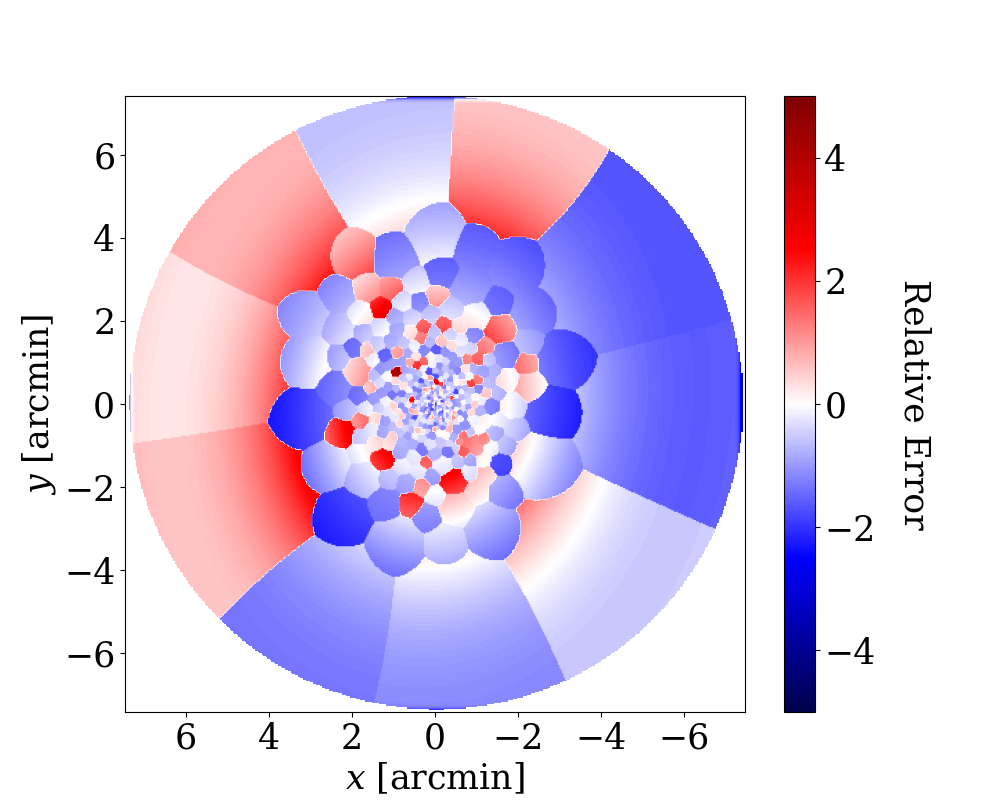}\hfill
    \\
    \includegraphics[width=0.33\textwidth]{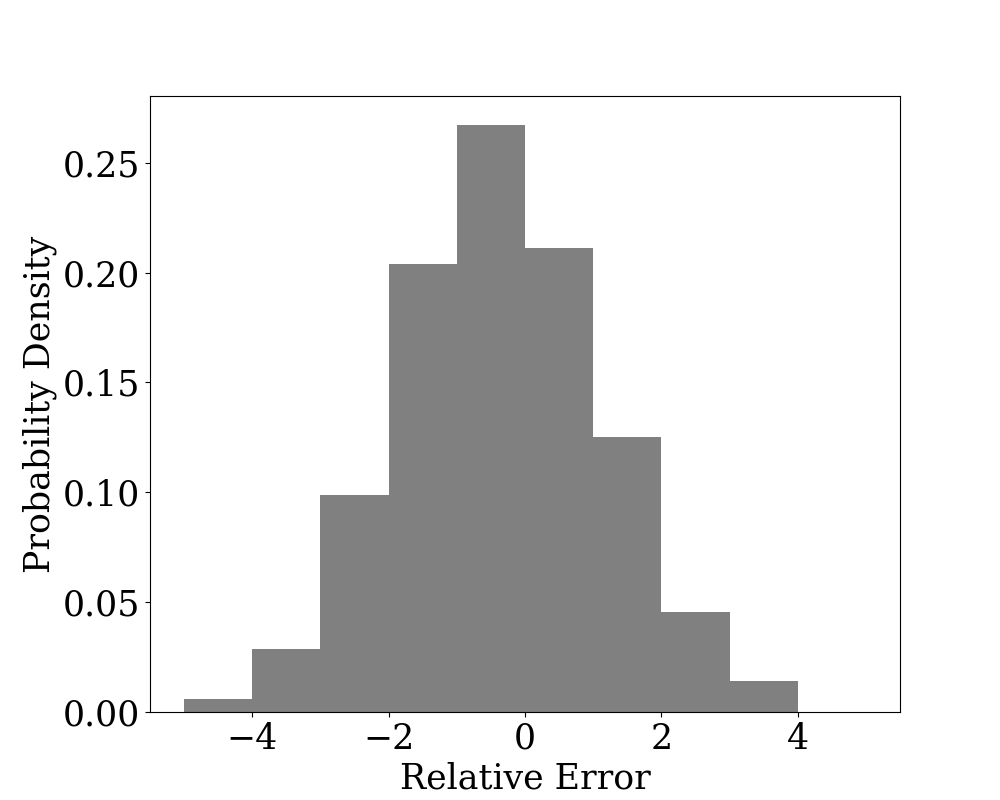}\hfill
    \includegraphics[width=0.33\textwidth]{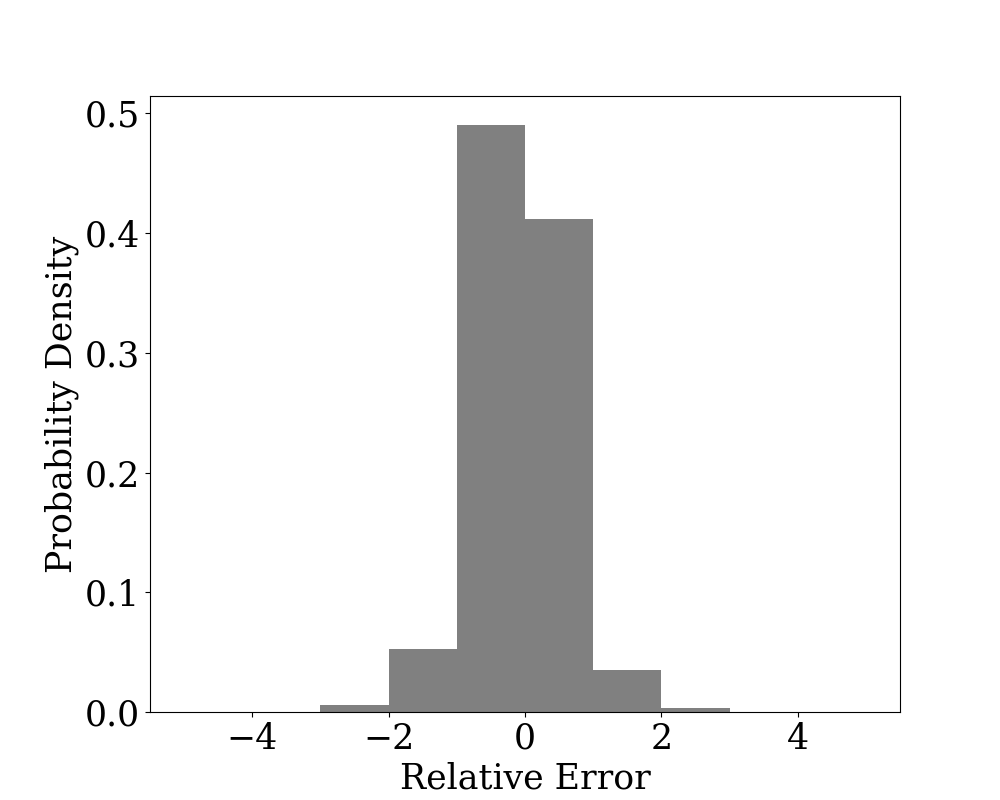}\hfill
    \includegraphics[width=0.33\textwidth]{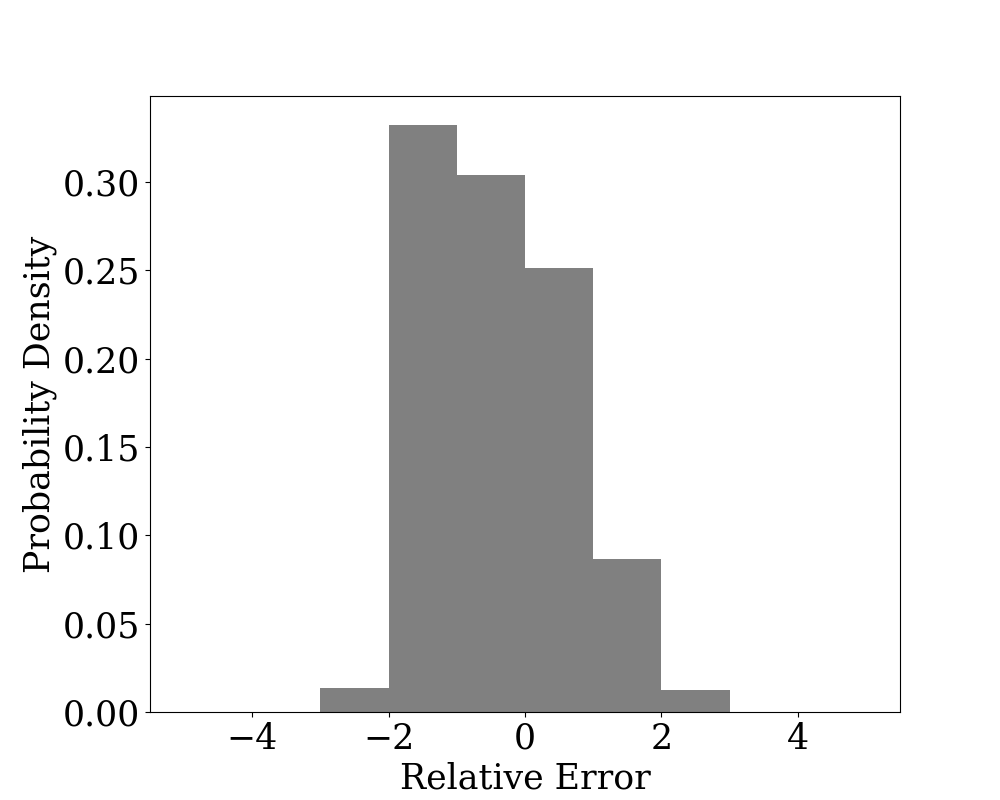}\hfill
    \caption{Reconstructed SZ and X-ray models of PSZ2 G313.33+61.13 generated using the recovered parameters from Fig.~\ref{fig:psz2g313-corner} ({\it top}). The difference between the observational data and the reconstructed model map above, in units of pixel-based error ({\it middle}). The histogram of the distribution of the relative error in the middle panels ({\it bottom}). The X-ray SB histogram takes into account both the residuals in the inner two-dimensional region, which includes 80\% of the emissions observed in the data, and the outer map region where we implemented one-dimensional analysis using azimuthal medians (see Fig.~\ref{fig:psz2g313-input}). In all cases, the residuals are distributed within $\pm 4\sigma$ level compared to the error. When the relative errors of the SZ, X-ray SB, and X-ray temperature are modeled with a Gaussian fit, their standard deviations align with $1.5\sigma$, $0.6\sigma$, and $1.1\sigma$ respectively. Comparatively, 63.8\%, 64.4\%, and 100\% of the data from SZ, X-ray SB, and X-ray temperature maps respectively exhibit pixel-based signal-to-noise ratios exceeding 1.5.}
    \label{fig:psz2g313-bestfit}
\end{figure*}

Figure~\ref{fig:psz2g313-corner} shows the posterior of the model parameters that describe our triaxial fit of PSZ2 G313.33+61.13, using the data from \Planck, ACT, and \xmm. We find axial ratios of $q_{\text{\tiny ICM,1}} = 0.65 \pm 0.02$ and $q_{\text{\tiny ICM,2}} = 0.79 \pm 0.02$. These values are consistent with previous results, but an order of magnitude more precise (Table~\ref{tab:psz2g313-summary}). Our fits indicate the major axis of Abell 1689 is almost perfectly aligned with the line of sight, with $\cos \theta \geq 0.96$ at 90\% confidence. While previous works also indicated such an alignment, a much wider range of orientations were allowed in those fits. We note that our analysis only includes statistical uncertainties on the fit, and the uncertainty due to data calibration is not taken into account here. Also, as the elongation parameter (Eq.~\ref{eq:elongation}), which is the ratio of the size of the ellipsoid along the observed line of sight to the major axis of the projected ellipse in the sky plane, quantifies the three-dimensional geometry of the triaxial ellipsoid model of the ICM. We thus present constraints on $e_\parallel$ rather than on $\varphi$ and $\psi$. The inferred $e_\parallel$ is well constrained in the fit to a value of $1.24 \pm 0.03$ and is consistent with the gas analysis result of \cite{2012MNRAS.419.2646S}, who found $e_\triangle = 0.66 \pm 0.21$, which corresponds to $1.15 \leq e_\parallel \leq 2.22$. Figure~\ref{fig:psz2g313-bestfit} shows the reconstructed SZ, X-ray SB and temperature maps of PSZ2 G313.33+61.13, incorporating the instrument response, generated using the recovered parameters from Fig.~\ref{fig:psz2g313-corner}. The difference map, which is created by taking the input data and subtracting the reconstructed model from it, reveals that the majority of the pixels exhibit relative errors that are spread within a range of $\pm 4\sigma$ (Fig.~\ref{fig:psz2g313-corner}). The residuals for the SZ, X-ray SB, and X-ray temperatures are distributed around zero. Their respective standard deviations are equivalent to $1.5\sigma$, $0.6\sigma$, and $1.1\sigma$ when fitted by a Gaussian.

For comparison, we performed an additional X-ray + SZ fit using only the \Planck\ SZ data, without incorporating the ground-based ACT data. We obtain posteriors that significantly deviate from our baseline fit with ACT data. We attributed this to the coarse angular resolution of \Planck\, which prevents it from resolving morphological features given the angular size of Abell 1689 at z=0.1832. To test this, we generated two sets of mock observations using the recovered parameters from our baseline fit to the observed data from both \Planck\ and ACT (along with \xmm). One mock was based on the properties of the $y$-map from \Planck\ + ACT, while the other mimicked the $y$-map with only \Planck\ SZ data, including the appropriate noise and PSF shape for each case. Our fit to the mock multiwavelength data with the \Planck\ + ACT $y$-map yields recovered parameters closely aligned with the input model, suggesting these data can accurately recover the input ICM shape. In contrast, the second mock observation based on the \Planck-only $y$-map produces a set of parameters significantly deviating from the input. This suggests that the SZ data from \Planck\ alone are insufficient to reliably fit our triaxial model, at least for a galaxy cluster with this specific shape at this specific redshift. This confirms that our fit to observed data using the \Planck-only $y$-map are likely biased. In a subsequent paper we will explore this issue in more detail, to better understand which types of galaxy clusters can (or cannot) be reliably reconstructed with the data available for CHEX-MATE.

Furthermore, in order to evaluate how the much higher overall signal to noise of the X-ray SB compared to the SZ and X-ray temperature impacts the results, we carried out an additional fit using the reduced $\chi^2$ for each of the three observables in order to weight them equally in the fit. The results of this fit indicate that there is only a minimal shift in the values of the derived geometric parameters based on this equal weighting of the three observables. Specifically, in the reduced $\chi^2$ fit, $q_{\text{\tiny ICM,1}}$ has a value of $0.70 \pm 0.04$, $q_{\text{\tiny ICM,2}}$ is $0.78 \pm 0.05$, and $e_\parallel$ stands at $1.22 \pm 0.07$. We also attempted to account for fluctuations in the calibration uncertainty, which can be especially important for the temperature profile \citep[e.g.,][]{2015A&A...575A..30S, 2022MNRAS.517.5594W}. We conducted model fits by introducing an additional $\sim10\%$ uncertainty of the temperature, but observed little changes in the parameters, with posteriors displaying similar levels of variation.

Table~\ref{tab:psz2g313-summary} presents a comparison of axial ratios, the (cosine of) inclination angle, and elongation parameter from our study. These metrics show notable improvement when compared to the gas analysis results reported by \cite{2012MNRAS.419.2646S}.
This marked improvement is attributable to multiple factors: our use of deeper new \xmm\ data not available to \cite{2012MNRAS.419.2646S}; our use of an \xmm\ SB image rather than a shallower {\it Chandra} SB image; our use of much higher quality SZ data from \Planck\ and ACT rather than from WMAP and SZA/OVRO/BIMA; and our improved analysis formalism making use of fully two-dimensional images for all of the observables rather than a projected elliptically averaged profile of X-ray SB and temperature along with a single-aperture photometric measurement of the SZ signal.
While a direct comparison between the current and previous ICM analysis of Abell 1689 is not straightforward due to data availability and differences in the ICM model, we performed an additional test. Instead of using the full two-dimensional images from the SZ and X-ray data, we processed the input data into one-dimensional projected profiles, which are more analogous to the data used in previous studies, and ran a model fit. The analysis yielded inferred geometrical parameters: $q_{\text{\tiny ICM,1}} = 0.61^{+0.14}_{-0.12}$, $q_{\text{\tiny ICM,2}} = 0.86^{+0.10}_{-0.14}$, $\cos \theta = 0.78^{+0.15}_{-0.39}$. These results align with the uncertainty levels found in the previous analysis, and suggest that much of the improvement in parameter constraints is related to the incorporation of higher quality, fully two-dimensional data available for our analysis.

As we will illustrate in subsequent studies, the derived geometric parameters of the ICM distribution, such as the elongation that quantifies the three-dimensional geometry, can be applied in conjunction with gravitational lensing measurements. For these fits, we will work under the assumption that the triaxial axes of the ICM and DM are co-aligned, but with axial ratios that are allowed to vary. The lensing analysis becomes crucial for discerning the triaxial shapes of DM, circumventing the need to rely on hydrostatic equilibrium or simulation-based corrections. Consequently, a comprehensive multi-probe analysis facilitates a characterization of the total matter distribution, which is essential for precise lensing-based mass calibrations \citep{2018ApJ...860L...4S}, along with allowing for a determination of the distribution of nonthermal pressure support \citep{2021MNRAS.505.4338S}.





\section{Conclusions}

We have improved a multi-probe analysis package to fit the three-dimensional ellipsoidal shapes of CHEX-MATE galaxy clusters. This package builds upon CLUMP-3D \citep{2017MNRAS.467.3801S}, which has been employed to analyze the triaxial shapes of CLASH clusters \citep{2018ApJ...860L...4S, 2018ApJ...860..126C, 2021MNRAS.505.4338S}. Specifically, we made the following improvements: (1) we model two-dimensional distributions of the SZ and X-ray temperature data, in contrast to the one-dimensional azimuthally averaged profiles in these quantities used by \cite{2017MNRAS.467.3801S}, (2) we parametrize electron density and pressure rather than density and temperature, reducing the number of parameters and speeding up the fit, and 3) we have ported the code to Python to facilitate a future public release. For the two-dimensional map analyses, we have added the capability to include publicly available SZ data from ground-based cosmic microwave background surveys such as that conducted with ACT, in addition to the default \Planck\ SZ maps.

We verified the triaxial analysis method through mock data analysis and applied it to the actual CHEX-MATE galaxy cluster, PSZ2 G313.33+61.13 (Abell 1689). The analysis effectively constrains the model geometry, in particular at the few percent level for the axial ratios. Our results are consistent with previous analyses of Abell 1689 available in the literature. Specifically, we find axial ratios of $q_{\text{\tiny ICM,1}}=0.65 \pm 0.02$ and $q_{\text{\tiny ICM,2}}=0.79 \pm 0.02$ and elongation parameter $e_\parallel = 1.24 \pm 0.03$. Compared to the similar gas-only analysis using X-ray and SZ data presented in \cite{2012MNRAS.419.2646S}, the axial ratios and elongation parameters in our study demonstrate a substantial improvement, with uncertainties an order of magnitude lower. Our results indicate that Abell 1689 has axial ratios typical of what is expected for the general population of galaxy clusters \citep{2002ApJ...574..538J, 2011ApJ...734...93L}, but a remarkably close alignment between the major axis and the line of sight. This alignment has resulted in exceptional lensing properties for Abell 1689, such as an abundance of strong lensing features \citep[e.g.,][]{2005ApJ...621...53B, 2007ApJ...668..643L}, one of the largest Einstein radii ever observed \citep[47\arcsec,][]{2010ApJ...723.1678C}, and an extremely large concentration of mass when fitted to a spherically symmetric model \citep[$c_\textrm{vir} = 12.8^{+3.1}_{-2.4}$ or $c_{200} = 10.2^{+2.6}_{-1.9}$,][]{2011ApJ...729..127U, 2020A&ARv..28....7U}. We thus conclude that there is nothing unusual about the triaxial shape of Abell 1689, other than its orientation. In addition, the estimated axial ratios of the cluster yield a triaxiality parameter $t=0.66$ \citep{1991ApJ...383..112F}. While the incorporation of lensing data is necessary for a direct quantitative comparison with DM axial ratios, the calculated $t$ classifies this halo as being close to the ``prolate'' population that comprises $\sim\!80$\% of the total cluster fraction in the DM-only simulations \citep{2017MNRAS.467.3226V}. The integration of lensing data for a comprehensive multiwavelength analysis, as well as the public release of the software and data products, will be addressed in subsequent papers of this series.


\begin{acknowledgements}
J.K. and J.S. were supported by NASA Astrophysics Data Analysis Program (ADAP) Grant 80NSSC21K1571. J.K. is supported by a Robert A. Millikan Fellowship from the California Institute of Technology (Caltech). J.K is supported by Korea Advanced Institute of Science \& Technology (KAIST) research fund for new faculty settlement. M.S. acknowledges financial contribution from contract ASI-INAF n.2017-14-H.0. and from contract INAF mainstream project 1.05.01.86.10. 
M.E.D. acknowledges partial support from the NASA ADAP, primary award to SAO with a subaward to MSU, SV9-89010.
S.E., F.G., and M.R. acknowledge the financial contribution from the contracts Prin-MUR 2022 supported by Next Generation EU (M4.C2.1.1, n.20227RNLY3 {\it The concordance cosmological model: stress-tests with galaxy clusters}), ASI-INAF Athena 2019-27-HH.0, ``Attivit\`a di Studio per la comunit\`a scientifica di Astrofisica delle Alte Energie e Fisica Astroparticellare'' (Accordo Attuativo ASI-INAF n. 2017-14-H.0), and from the European Union’s Horizon 2020 Programme under the AHEAD2020 project (grant agreement n. 871158). This research was supported by the International Space Science Institute (ISSI) in Bern, through ISSI International Team project \#565 ({\it Multi-Wavelength Studies of the Culmination of Structure Formation in the Universe}).
A.I., E.P., and G.W.P. acknowledge support from CNES, the French space agency.
K.U. acknowledges support from the National Science and Technology Council of Taiwan (grant 109-2112-M-001-018-MY3) and from the Academia Sinica (grants AS-IA-107-M01 and AS-IA-112-M04).
B.J.M. acknowledges support from STFC grant ST/V000454/1. L.L. acknowledges financial contribution from the INAF grant 1.05.12.04.01.
\end{acknowledgements}

\bibliographystyle{aa} 
\bibliography{bib} 

\end{document}